\documentclass[fleqn,usenatbib]{mnras}

\usepackage{newtxtext,newtxmath}

\usepackage[T1]{fontenc}

\DeclareRobustCommand{\VAN}[3]{#2}
\let\VANthebibliography\thebibliography
\def\thebibliography{\DeclareRobustCommand{\VAN}[3]{##3}\VANthebibliography}

\usepackage{graphicx}	
\usepackage{amsmath}	

\usepackage[dvipsnames]{xcolor}

\title[Baryons in SIDM mergers]{The role of baryons in self-interacting dark matter mergers}

\author[M. S. Fischer et al.]{Moritz S. Fischer,$^{1,2,3}$\thanks{E-mail: mfischer@usm.lmu.de (LMU)}
Nils-Henrik Durke,$^{3}$\thanks{Deceased}
Katharina Hollingshausen,$^{4,6}$
Claudius Hammer,$^{5,6}$
\newauthor{Marcus Br\"{u}ggen,$^{3}$
Klaus Dolag$^{1,7}$}
\\
$^{1}$Fakult\"at für Physik, Universit\"ats-Sternwarte, Ludwig-Maximilians-Universit\"at M\"unchen, Scheinerstr. 1, D-81679 M\"unchen, Germany\\
$^{2}$Excellence Cluster ORIGINS, Boltzmannstrasse 2, D-85748 Garching, Germany\\\
$^{3}$Hamburger Sternwarte, Universit\"at Hamburg, Gojenbergsweg 112, D-21029 Hamburg, Germany\\
$^{4}$Fakult\"at für Mathematik und Informatik, Georg-August-Universität G\"ottingen, Bunsenstr. 3-5, D-37073 G\"ottingen, Germany\\
$^{5}$Johann-Wolfgang-Goethe-Gymnasium, August-Keiler-Straße 34, D-76726 Germersheim, Germany\\
$^{6}$Heidelberger Life-Science Lab, DKFZ, Technologiepark 4, Im Neuenheimer Feld 581, D-69120 Heidelberg, Germany\\
$^{7}$Max-Planck-Institut f\"ur Astrophysik, Karl-Schwarzschild-Str. 1, D-85748 Garching, Germany\\
}

\date{Accepted XXX. Received YYY; in original form ZZZ}

\pubyear{2023}

\begin{document}
\label{firstpage}
\pagerange{\pageref{firstpage}--\pageref{lastpage}}
\maketitle

\begin{abstract}
Mergers of galaxy clusters are promising probes of dark matter (DM) physics. For example, an offset between the DM component and the galaxy distribution can constrain DM self-interactions.
We investigate the role of the intracluster medium (ICM) and its influence on DM--galaxy offsets in self-interacting dark matter models.
To this end, we employ Smoothed Particle Hydrodynamics + $N$-body simulations to study idealized setups of equal- and unequal-mass mergers with head-on collisions.
Our simulations show that the ICM hardly affects the offsets arising shortly after the first pericentre passage compared to DM-only simulations. But later on, e.g.\ at the first apocentre, the offsets can be amplified by the presence of the ICM.
Furthermore, we find that cross-sections small enough not to be excluded by measurements of the core sizes of relaxed galaxy clusters have a chance to produce observable offsets.
We found that different DM models affect the DM distribution and also the galaxy and ICM distribution, including its temperature.
Potentially, the position of the shock fronts, combined with the brightest cluster galaxies, provides further clues to the properties of DM.
Overall our results demonstrate that mergers of galaxy clusters at stages about the first apocentre passage could be more interesting in terms of DM physics than those shortly after the first pericentre passage.
This may motivate further studies of mergers at later evolutionary stages.
\end{abstract}




\section{Introduction} \label{sec:introduction}

Galaxy clusters allow us to learn about dark matter (DM) and constrain its nature in various manners.
Particularly interesting are mergers of galaxy clusters.
It turned out that their phenomenology is for DM alternatives quite challenging to explain.
This concerns the distribution of luminous matter compared to the gravitational lensing signal, which does not match the former one \citep[e.g.][]{Clowe_2006}.
In consequence, DM models appear to be more promising compared to theories such as modified Newtonian dynamics.
Furthermore, mergers could potentially probe whether DM is collisionless as in the cosmological standard model, Lambda cold dark matter ($\Lambda$CDM) or is subject to self-interactions.
Self-interacting dark matter (SIDM) models could lead to offsets between the DM and the galaxies hosted by the clusters.
This feature is often though to be unique for SIDM, but in principle other DM candidates could also lead to such offsets \citep[e.g.][]{Bak_2021}.

Initially, DM self-interactions have been proposed by \cite{Spergel_2000} to solve issues of $\Lambda$CDM on small scales, such as the core-cusp problem \citep[for a review on the small-scale crisis see][]{Bullock_2017}.
As self-interactions transfer heat to the centre of the halo they can lower the central density, i.e.\ create DM cores.
This has been shown, for example, by \cite{Dave_2001}.
SIDM has been found to be promising to solve other small-scale issues too.
It can create diverse rotation curves
\citep[e.g.][]{Creasey_2017, Kamada_2017, Robertson_2018} and may be able to solve the too-big-to-fail
problem \citep[e.g.][]{Zavala_2013, Elbert_2015, Kaplinghat_2019}.
Further details about SIDM can be found in one of the review articles \citep[][]{Tulin_2018, Adhikari_2022}.

The first studies of SIDM were limited to isotropic cross-sections and mostly studied velocity-independent self-interactions \citep[e.g.][]{Burkert_2000}.
There exist a huge variety of SIDM models which have a diverse phenomenology.
All the qualitative differences between the DM candidates have not yet been fully explored.
There are numerous studies that have taken velocity-dependent self-interactions into account \citep[e.g.][]{Colin_2002, Colin_2003, DOnghia_2003}.
In the context of mergers, this has been done by \cite{Robertson_2017b}.
This is well motivated from a particle physics point of view \citep[e.g.][]{Ackermann_2009, Buckley_2010, Loeb_2011, van_den_Aarssen_2012, Tulin_2013a} as well as from an astrophysical one \citep[e.g.][]{Kaplinghat_2016}.
Furthermore, several studies investigated anisotropic cross-sections \citep[e.g.][]{Kahlhoefer_2014, Kahlhoefer_2015, Kummer_2018, Kummer_2019, Robertson_2017b, Banerjee_2020, Nadler_2020, Fischer_2021a, Fischer_2021b, Fischer_2022}.
Various DM candidates can produce an anisotropic cross-section, as for example mirror DM \citep{Blinnikov_1983, Kolb_1985, Berezhiani_1996, Foot_2004}, atomic DM \citep{Kaplan_2010, Cline_2012, Cyr-Racine_2013} and some other hidden sector DM models \citep{Feng_2009, Foot_2015, Boddy_2016}.
Some work has been done on dissipative or inelastic self-interactions including multi-state scattering \citep[e.g.][]{Schutz_2015, Essig_2019, Vogelsberger_2019, Chua_2020, Huo_2020, ONeil_2023}.
Inelastic scattering occurs in candidates such as atomic DM \citep{Cline_2014, Boddy_2016} or strongly interacting composite DM \citep{Boddy_2014}.

In this paper, we focus on mergers of galaxy clusters.
They have mainly been studied with CDM, especially dissociative mergers, where the ICM is separated from the DM haloes.
In consequence, the X-ray emission does not coincide with the distribution of the cluster galaxies and the majority of the mass as inferred from gravitational lensing.
Known examples for such systems include the Bullet Cluster \citep[e.g.][]{Springel_2007, Mastropietro_2008, Lage_2014}, the `El Gordo' cluster \citep[e.g.][]{Donnert_2014, Molnar_2015, Zhang_2015, Kim_2021}, the `Sausage' cluster \citep[e.g.][]{Donnert_2017, Molnar_2017}, A1758N \citep{Machado_2015b, Monteiro-Oliveira_2017}, and ZwCl008.8+52 \citep{Molnar_2018}.
Within the cold collisionless DM paradigm, one would expect the distribution of galaxies to coincide with the DM. An offset between the two would be a strong hint to new DM physics and various authors have tried to find such offsets. However, claims of observed large DM--galaxy offsets could not withhold a more detailed analysis \citep[e.g.][]{Bradac_2008, Dawson_2012, Dawson_2013, Jee_2014, Jee_2015, Harvey_2017, Peel_2017, Taylor_2017, Wittman_2018}.
To date, there is no conclusive evidence from observations for large offsets between the DM and galaxy distribution.

This implies constraints on models of SIDM as self-interactions provide a mechanism to create such offsets.
The scattering of the DM particles can lead to an effective drag force that decelerates the DM component but does not directly affect the galaxies.
The first simulations of SIDM mergers employed an isotropic cross-section \citep[e.g.][]{Randall_2008, Kim_2017b, Robertson_2017a} and soon also an anisotropic cross-section \citep{Robertson_2017b}.
In the context of mergers, very anisotropic cross-sections that typically scatter about a tiny angle may be more interesting, because they can give rise to a strong drag force and thus lead to much larger offsets than isotropic cross-sections do \citep{Kahlhoefer_2014, Fischer_2021a, Fischer_2021b}.
If particles scatter by a large angle, they transfer more momentum per scattering event compared to a small scattering angle.
Thus fewer scattering events are required to alter the distribution of DM significantly.
We call these self-interactions rare, in contrast to frequent self-interactions, which we use to refer to cross-sections with a tiny scattering angle.
We mean by frequently self-interacting dark matter (fSIDM) the limit in which the momentum transfer cross-section stays constant but the typical scattering angle approaches zero.
The momentum transfer cross-section is given as,\footnote{Assuming identical particles, the definition is equivalent to the one recommended by \cite{Robertson_2017b} and \cite{Kahlhoefer_2017}.}
\begin{equation} \label{eq:momentum_transfer_cross_section}
    \sigma_\mathrm{\tilde{T}} = 4\pi \int_0^1 \frac{\mathrm{d} \sigma}{\mathrm{d} \Omega_\text{cms}} (1 - \cos \theta_\text{cms}) \mathrm{d} \cos \theta_\text{cms} \, .
\end{equation}
Where $(1 - \cos\theta)$ gives the relative change of momentum in the longitudinal direction.
As the momentum transfer cross-section more strongly weights interactions with large momentum transfer than those with little momentum transfer, it provides a better approximation for how strong self-interactions impact the DM distribution compared to the total cross-section.
Moreover, \cite{Fischer_2022} found that the momentum transfer cross-section roughly matches fSIDM and isotropic scattering in terms of the DM halo density profiles.
In line with previous work we use the momentum transfer cross-section to match the different angular dependencies \citep[][]{Robertson_2017b, Fischer_2021a, Fischer_2021b}.
We note, that also different matching procedures have been proposed, in particular the viscosity cross-section \citep[e.g.][]{Tulin_2013a, Colquhoun_2021, Yang_2022D}.

Previous theoretical merger studies with rSIDM did not find large offsets \citep[][]{Randall_2008, Kim_2017b, Robertson_2017a, Robertson_2017b} for viable cross-sections.
However, most of these studies focused on mergers shortly after their first pericentre passage, with the exception of \cite{Kim_2017b}.
In contrast, recent studies \citep{Fischer_2021a, Fischer_2021b} pointed out that later merger stages about the first apocentre passage might be more interesting.
Moreover, the difference between rSIDM and fSIDM when comparing the same momentum transfer cross-section can be remarkable.
While rSIDM does not show any significant offsets, fSIDM can give offsets larger than 50 kpc for a galaxy cluster merger \citep{Fischer_2021b}.
In consequence, mergers of galaxy clusters are an interesting test bed for DM self-interaction, in particular small-angle scattering.

Although the possibility to probe the nature of DM by DM--galaxy offsets has fostered substantial work, our theoretical understanding of mergers is still incomplete.
This includes the role of the ICM at late merger stages and in fSIDM mergers.
Due to the non-observation of significant DM--galaxy offsets, the motivation to pursue such studies may have decreased.
Nevertheless, mergers of galaxy clusters are suitable systems to constrain the nature of DM.
Especially for fSIDM, vanishing offsets could imply comparable or even more stringent bounds on the cross-section compared to other observables such as the core size of galaxy clusters \citep[e.g.][]{Andrade_2021, Sagunski_2021, Eckert_2022}.

Moreover, a potential DM--galaxy offset may not be the only observable that could help to falsify DM models with mergers of galaxy clusters.
By \cite{Fischer_2021b}, it was found that the morphology of the galaxy distribution is indirectly affected by DM self-interactions, too.
Within CDM, the morphology of the ICM has been studied in more detail \citep[e.g.][]{Keshet_2021, ZuHone_2022}, but for alternative DM models, there is very little work.

In this paper, we investigate the influence of the ICM in head-on collisions of galaxy clusters with isotropic and small-angle scattering.
To this end, we perform $N$-body simulation of equal- and unequal-mass mergers and study their evolution.

In Section~\ref{sec:methods} we describe our numerical setup, including the initial conditions (ICs) and numerical parameters, as well as the methods we use to analyse the simulation data.
A presentation of the results, including DM--galaxy offsets, follows in Section~\ref{sec:results}.
In Section~\ref{sec:discussion} we discuss the shortcomings of this work as well as directions for further research.
Finally, we summarize and conclude in Section~\ref{sec:conclusion}.
Additional information on the initial conditions and the gravitational potential is provided in the appendices.

\section{Numerical setup and methods} \label{sec:methods}

In this section, we first describe the ICs and the simulation parameters.
Subsequently, we explain how we analyse the simulation data.

\subsection{Simulations}

\begin{table*}
    \centering
    \begin{tabular}{c|c|c|c|c|c|c}
        $M_\mathrm{vir}$ & $r_s$ & $\rho_{0,\mathrm{DM}}$ & $\rho_{0,\mathrm{Gal}}$ & $r_b$ & $\rho_b$ & N\\
        $(\mathrm{M_\odot})$ & (kpc) & ($\mathrm{M_\odot}$ kpc$^{-3}$) & ($\mathrm{M_\odot}$ kpc$^{-3}$) & (kpc) & ($\mathrm{M_\odot}$ kpc$^{-3}$) & \\ \hline
        $10^{15}$ & 389.31 & $1.14 \times 10^6$ & $2.28 \times 10^4$ & 194.78 & $9.72 \times 10^5$ & $3\times8\,806\,762 + 1$\\
        $2 \times 10^{14}$ & 194.76 & $1.63 \times 10^6$ & $3.27 \times 10^4$ & 97.38 & $1.4 \times 10^6$ & $3\times1\,583\,425 + 1$\\
        $10^{14}$ & 144.53 & $1.89 \times 10^6$ & $3.79 \times 10^4$ & 72.16 & $1.64 \times 10^6$ & $3\times757\,593 + 1$
    \end{tabular}
    \caption{The table gives the parameter of the haloes we use for our ICs. First, we give the virial mass, $M_\mathrm{vir}$, followed by the NFW parameters for the DM and galaxy components. The scale radius, $r_s$, is the same for both components, but they differ in the density parameter, $\rho_0 \equiv 4 \, \rho(r_s)$. Furthermore, we give parameters of the Burkert profile which describes the ICM. $r_b$ is the radial parameter and $\rho_b$ is the density parameter.
    The last column gives the number of particles per halo, as the sum of all components.}
    \label{tab:halos}
\end{table*}

The ICs for the merger simulations consist of two DM haloes.
Each halo is composed of three components, DM, galaxies and the ICM.
The density of the DM and the galaxies follow a Navarro--Frank--White (NFW) profile \citep{Navarro_1996}:
\begin{equation} \label{eq:NFW}
    \rho_\mathrm{NFW}(r) = \frac{\rho_0}{\frac{r}{r_s} \left( 1 + \frac{r}{r_s} \right)^2 } \,.
\end{equation}
For the density of the ICM we use a Burkert profile \citep{Burkert_1995}:
\begin{equation} \label{eq:Burkert}
    \rho_\mathrm{Burkert}(r) = \frac{\rho_b \, r_b^3}{(r + r_b) \, (r^2 + r_b^2)} \,.
\end{equation}
The detailed parameters for the DM haloes are given in Tab.~\ref{tab:halos}.
As the total mass of an NFW profile is infinite, it needs to be truncated.
We do so at 20 times the scale radius, $r_s$.
The ICs are generated by drawing random numbers from the probability density function that correspond to the analytic density profile, i.e.\ for DM and galaxies to the NFW profile (Eq.~\ref{eq:NFW}) and for the ICM to the Burkert profile (Eq.~\ref{eq:Burkert}).
This allows us to set the positions of the numerical particles.
Integrating the Jeans Equation for this particle distribution, including all components, yields the particle velocities and the ICM temperature.

The numerical particle masses are for DM, $m_\mathrm{DM} = 2 \times 10^8 \, \mathrm{M_\odot}$, and for galaxies, $m_\mathrm{Gal} = 4 \times 10^6 \, \mathrm{M_\odot}$, and for the ICM particles, $m_\mathrm{ICM} = 3 \times 10^7 \, \mathrm{M_\odot}$.
We choose the same number of particles for each of these three components to have a similar error in the peak identification (see sec.~\ref{sec:analysis}).
In addition, we place one more massive particle at the centre of each halo to mimic the brightest cluster galaxy (BCG).
The BCG particle has a mass of $m_\mathrm{BCG} = 7 \times 10^{10} \, \mathrm{M_\odot}$, which is much lower than typical BCG masses.
This choice is motivated by reducing numerical artefacts. With our mass choice, we follow \cite{Kim_2017b}.
Overall our galaxy cluster consist of roughly $\approx 85.5\%$ DM,  $\approx 12.8\%$ ICM, and $\approx 1.7\%$ galaxies.
We combine those haloes to simulate mergers with different merger mass ratios (MMR).
For all of them, the impact parameter vanishes, $b=0$.
Further details are given in Tab.~\ref{tab:runs}.

We run the simulations using the cosmological $N$-body code \textsc{gadget-3}, with an updated version of the SPH formulation \citep{2016MNRAS.455.2110B} and with collisionless DM as well as collisional DM.
Precisely speaking, we use \textsc{OpenGadget3}, which is an upgraded version of \textsc{gadget-2} \citep{Springel_2005}. It contains various additional modules such as cosmic rays \citep{Boess_2022} or meshless finite mass hydrodynamics \citep{Groth_2023} to name the most recent ones.

For the DM self-interactions we use the implementation described by \citet{Fischer_2021a, Fischer_2021b}.
We simulate frequent self-interactions as well as rare self-interactions using an isotropic cross-section.
For the latter one, we apply the relabelling of the scattering particles if the scattering angle is larger than 90° as previously described in appendix~A of \cite{Fischer_2021b}.
The purpose of the relabelling is to avoid particles changing the halo to which they belong without having any physical effect as they are identical.
The gravitational softening length is set to a constant value of $\epsilon = 1.2 \, \mathrm{kpc}$ of every component.
For the SIDM implementation, we use an adaptive kernel size determined by the distance to the neighbouring particles.
The number of neighbours within the kernel is set to $N_\mathrm{ngb} = 64$. 
For the kernel function, we employ a spline kernel \citep{Monaghan_1985} as previously done by \cite{Fischer_2021b}.
To simulate the ICM we use smoothed-particle hydrodynamics \cite[SPH,][]{Gingold_1977}.
We use an updated SPH formulation \citep{2016MNRAS.455.2110B}, where we have chosen to use a Wendland $C^6$ kernel \citep{Dehnen_2012} with $N_\mathrm{ngb} = 296$ and an updated viscosity scheme which especially allows resolving the turbulent motion within the ICM \citep{2005MNRAS.364..753D, 2016MNRAS.455.2110B}.
The minimum temperature is set to 10 Kelvin \citep[for more details see][]{Springel_2005}.
We do not employ any additional physics for the ICM. In particular, we do not model any cooling or heating of the ICM.

\begin{table}
    \centering
    \begin{tabular}{c|c|c|c|c|c|c}
        $M_\mathrm{vir, main}$ & MMR & $d_\mathrm{ini}$& $\Delta v_\mathrm{ini}$ & $\sigma_\mathrm{\tilde{T}}/m$\\
        $(\mathrm{M_\odot})$ & & (kpc) & (km s$^{-1}$) & $(\mathrm{cm}^2 \, \mathrm{g}^{-1})$ \\ \hline
        $10^{15}$ & 1:1 & 4000 & 1000 & 0.0, 0.1, 0.3, 0.5 \\
        $10^{15}$ & 1:5 & 4000 & 1000 & 0.0, 0.1, 0.3, 0.5 \\
        $10^{15}$ & 1:10 & 4000 & 1000 & 0.0, 0.1, 0.3, 0.5 \\
    \end{tabular}
    \caption{The initial condition and simulation parameters are given for the runs presented in this paper.
    The first column gives the virial mass, $M_\mathrm{vir, main}$, of the main halo.
    Next, MMR denotes the merger mass ratio in terms of the virial mass.
    Initially, the two halo centres are separated by $d_\mathrm{ini}$, their initial relative velocity is $\Delta v_\mathrm{ini}$ and they are all on a  head-on collision trajectory.
    $\sigma_\mathrm{\tilde{T}} / m$ gives the self-interaction cross-section (see equation~\ref{eq:momentum_transfer_cross_section}) and the values have been simulated with rare and frequent self-interactions, apart from $\sigma_\mathrm{\tilde{T}} / m =0.0$ which corresponds to CDM.
    }
    \label{tab:runs}
\end{table}

\subsection{Analysis} \label{sec:analysis}

We use two methods to identify the peak of the matter components.
The first is based on the gravitational potential and the second is based on isodensity contours.
For the first one, we compute the peak of each halo and component by only considering the particles that initially belong to the halo. We compute their gravitational potential and search for the minimum.
This implies that the peak search is performed in three dimensions.\\

For the isodensity-based peaks, we always use all particles that belong to the respective component, e.g.\ DM or galaxies. As in observational analyses, we do not require knowledge to which halo a particle initially belongs.
Moreover, the peaks are searched in the 2d-projected density.
However, is not obvious how to assign the peaks to haloes, making it hard to trace a halo over time.
Hence, we linearly extrapolate the peak position of a halo from the past and compare it to
the current peaks, associating the peak which is closest to the prediction with the halo.
In order to estimate the error on the peak position we bootstrap the distribution and determine the peak again. This routine is performed 24 times.

Further details about the peak finding are explained by \cite{Fischer_2021b}, we use the same implementation here.

There are multiple ways to define the offset between the DM and galactic component.
They differ in the way the sign of the offset is determined.
Again, we follow the approach by \cite{Fischer_2021b},
\begin{equation} \label{eq:offset}
    \mathrm{offset} \equiv x_\mathrm{DM} - x_\mathrm{gal} \,.
\end{equation}
Here, $x$ denotes the spatial coordinate of the peak along the merger axis.

In addition to the peak positions of the matter components, we also measure the location of the shock front.
Therefore we calculate the temperature gradient along the merger axis in three dimensions.
The steepest gradient then defines the shock position, which specifies the intersection of the merger axis and the shock front.

\begin{figure*}
    \centering
    \includegraphics[width=\textwidth]{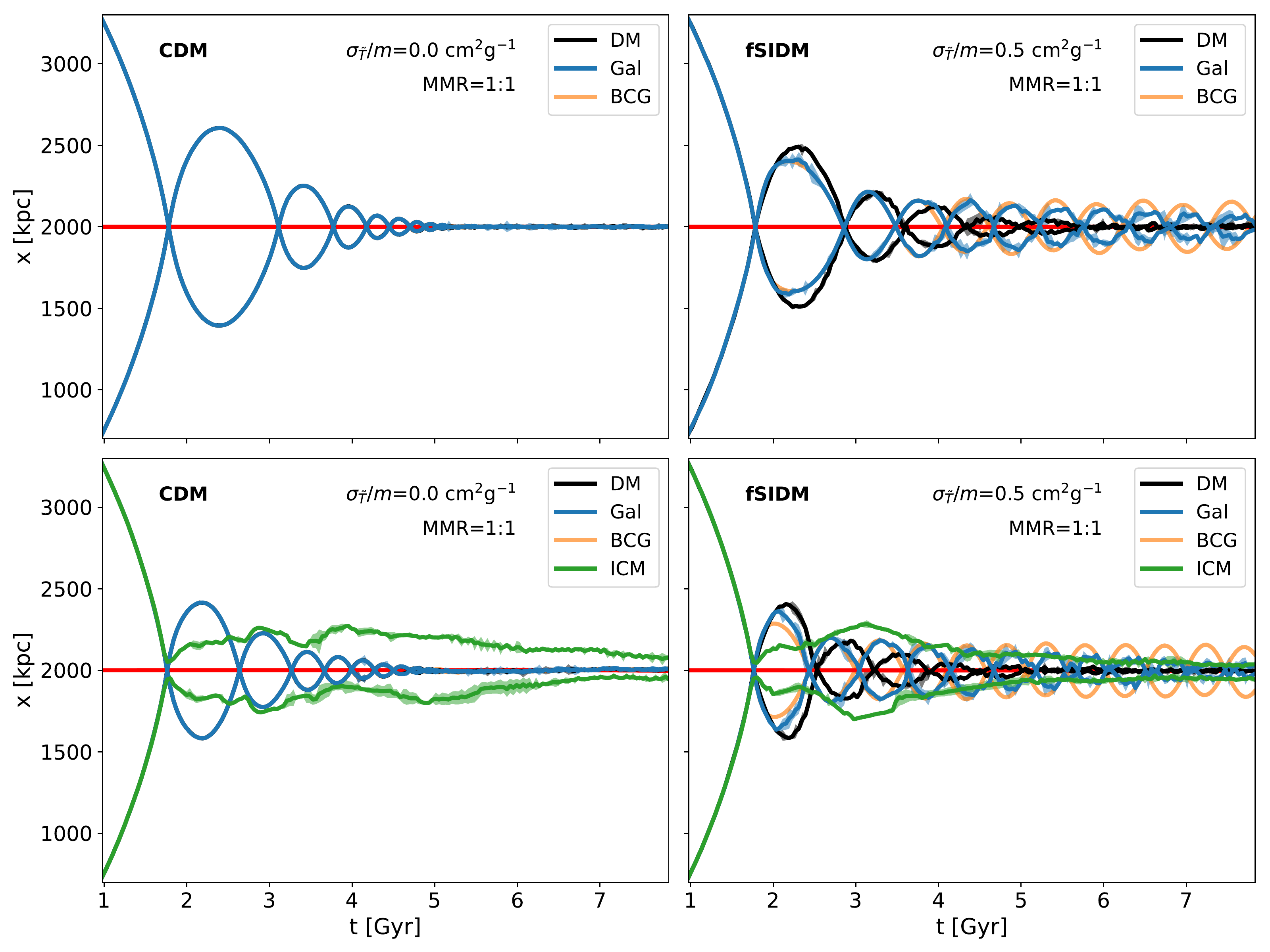}
    \caption{The peak positions are shown as function of time for our equal-mass mergers evolved with CDM (let-hand side) and fSIDM (right-hand side) with $\sigma_\mathrm{\Tilde{T}}/m = 0.5 \, \mathrm{cm}^2 \, \mathrm{g}^{-1}$.
    The components are indicated in different colours, DM in black, galaxies in blue, BCGs in faint orange, and ICM in green.
    We indicate the centre of mass with a red line.
    The upper panels are produced from the DMO simulations of \protect\cite{Fischer_2021b} with the right-hand panel being a reprint of \citet[][fig.~10, upper panel]{Fischer_2021b}.
    The bottom row gives the results of our new simulations including the ICM.
    }
    \label{fig:peakpos_equal}
\end{figure*}

In order to compare simulations with different cross-sections in time, we introduce the internal merger time $\tau$ as previously by \cite{Fischer_2021b}:
\begin{equation} \label{eq:merger_time}
    \tau \equiv \frac{t - t_\mathrm{first\,pericentre}}{t_m} , 
\end{equation}
where $t_m = t_\mathrm{second\,pericentre} - t_\mathrm{first\,pericentre}$ denotes the merger time.
By definition, $\tau=0$ corresponds to the first pericentre passage and $\tau=1$ to the second pericentre passage.
In practice, we compute the time $\tau$ using the BCGs which makes $\tau$ independent of the peak finding algorithm and thus always known.

\section{Results} \label{sec:results}

Here, we first present the evolution of the peak positions of the individual components and haloes.
We compare DMO and ICM simulations and describe the effect of DM self-interactions on the merger evolution.
Then, we compute the offsets and investigate the influence of the ICM before focusing on the core sloshing in the late merger stages. Finally, we take a look at the merger morphology in terms of density, temperature and investigate the propagation of the shock fronts.

\subsection{Peak positions}

\begin{figure*}
    \centering
    \includegraphics[width=\textwidth]{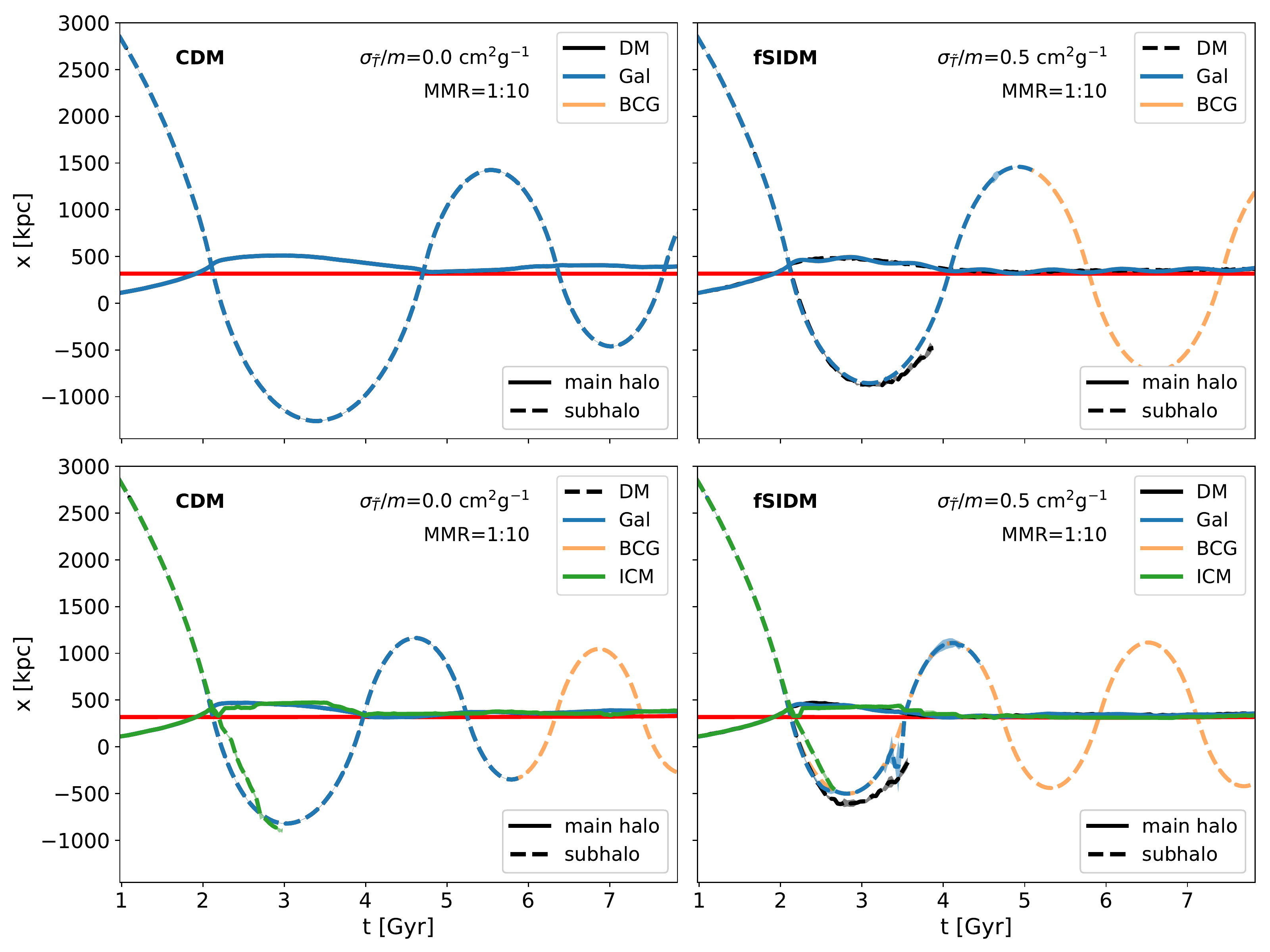}
    \caption{We show the peak position as a function of time for our 1:10 merger.
    This is the same as in Fig.~\ref{fig:peakpos_equal} but for a different MMR.
    The top row for CDM is a reprint of fig.~5 of \protect\cite{Fischer_2021b}.
    We note that peak finding is more difficult for unequal-mass mergers.
    For late times when the peak positions become very inaccurate, we do not show them.
    }
    \label{fig:peakpos_unequal}
\end{figure*}

To study the role of DM self-interactions and the ICM we plot the peak positions obtained with the potential-based peak finding method as a function of time.
We first begin with our equal-mass mergers in Fig.~\ref{fig:peakpos_equal}.
In the upper panels, we show the DMO results and in the lower panels, we show the ICM simulations.
We find that the peak positions are oscillating about the centre of mass (red line) and coalesce over time.
For the CDM simulations, the amplitude is vanishing at late times.
As DM and galaxies are represented by collisionless particles, their peaks behave the same.
When including the ICM (bottom-left panel) the merger amplitude shrinks and the merger time, too, i.e.\ the haloes coalesce earlier.
The time before the first pericentre passage is hardly affected, only the later evolution differs between the simulations.
The ICM behaves strongly collisionally, as a consequence, the ICM peaks do not cross the centre of mass.
Instead, the ICM is strongly decelerated and heats up.
In consequence, it is expanding and as the overall potential gradients become flatter, the ICM peaks move further outside.
The position of the ICM peaks could be strongly influenced by shock waves.
Over time the gas mixes and the distributions of the two haloes become more similar.
Consequently, the ICM peaks move closer to the centre of mass.
We show the evolution of the ICM in more detail in Section~\ref{sec:morphology} and \ref{sec:shock}.

Moreover, due to the ICM, there is more mass close to the barycentre compared to the CDM simulations.
This deepens the gravitational potential and can explain the much smaller apocentre distances of the DM and the galaxies.
We show this in greater detail in Appendix~\ref{sec:grav_pot}.
Furthermore, the ICM shows shallow density gradients, which lead to large errors in the peak positions.
As the indicated evolution of the two ICM peaks is not very symmetric, it becomes obvious that the error is larger than indicated in the plot.
In consequence, the applied bootstrapping method underestimates the error.

When turning frequent self-interactions (right-hand side) on, the DM component behaves differently from the galaxies.
These differences arise at the first pericentre passage.
Here, the DM component is decelerated by the effective drag force arising from the DM self-interactions.
In consequence, the DM peaks are closer to the barycentre.
This changes on the way to the first apocentre passage, the DM peaks pass the galaxy peaks and are located at a larger distance.
Moreover, the apocentre distance of the DM peaks is larger than the apocentre distance of the galactic peaks.
This may seem surprising, but can be understood in terms of the evolution of the gravitational potential gradients, which at this stage become flatter with time.
At about the first apocentre passage, quite large offsets are arising between the DM and the galactic component.
We note that the offsets between the BCGs and the DM are even larger.
In Section~\ref{sec:offsets} we study the offsets and the influence of the ICM on them in larger detail.
At late merger times, when the DM haloes have coalesced, the galaxies and BCGs keep oscillating.
This is in contrast to the CDM simulations, we investigate this aspect further in Section~\ref{sec:core_sloshing}.

In addition to the equal-mass mergers, we studied collisions with unequal mass ratios.
In Fig.~\ref{fig:peakpos_unequal} we show the results for the 1:10 merger, with CDM and fSIDM employing a cross-section of $\sigma_\mathrm{\Tilde{T}}/m = 0.5 \, \mathrm{cm}^2 \, \mathrm{g}^{-1}$.
The ICM has a similar effect as in the equal-mass mergers.
It leads to a shorter merger time and smaller amplitude.
In contrast to the equal-mass mergers, the ICM of the subhalo penetrates the main halo and the peak has a trajectory that is more similar to the DM or galactic component than for the equal-mass mergers.
The self-interactions also shorten the merger time and decrease the amplitude, similar to the equal-mass mergers.
This is not only the case for the DMO runs but is still present in the simulations including the ICM.
In the upper right-hand side panel for the DMO fSIDM simulation, an offset between the DM and galaxies is visible.
Interestingly, this offset becomes much larger when including the ICM, see lower right-hand side panel.
Next, we take a closer look at the offsets and the role of the ICM for them.

\begin{figure*}
     \centering
        \includegraphics[width=\textwidth]{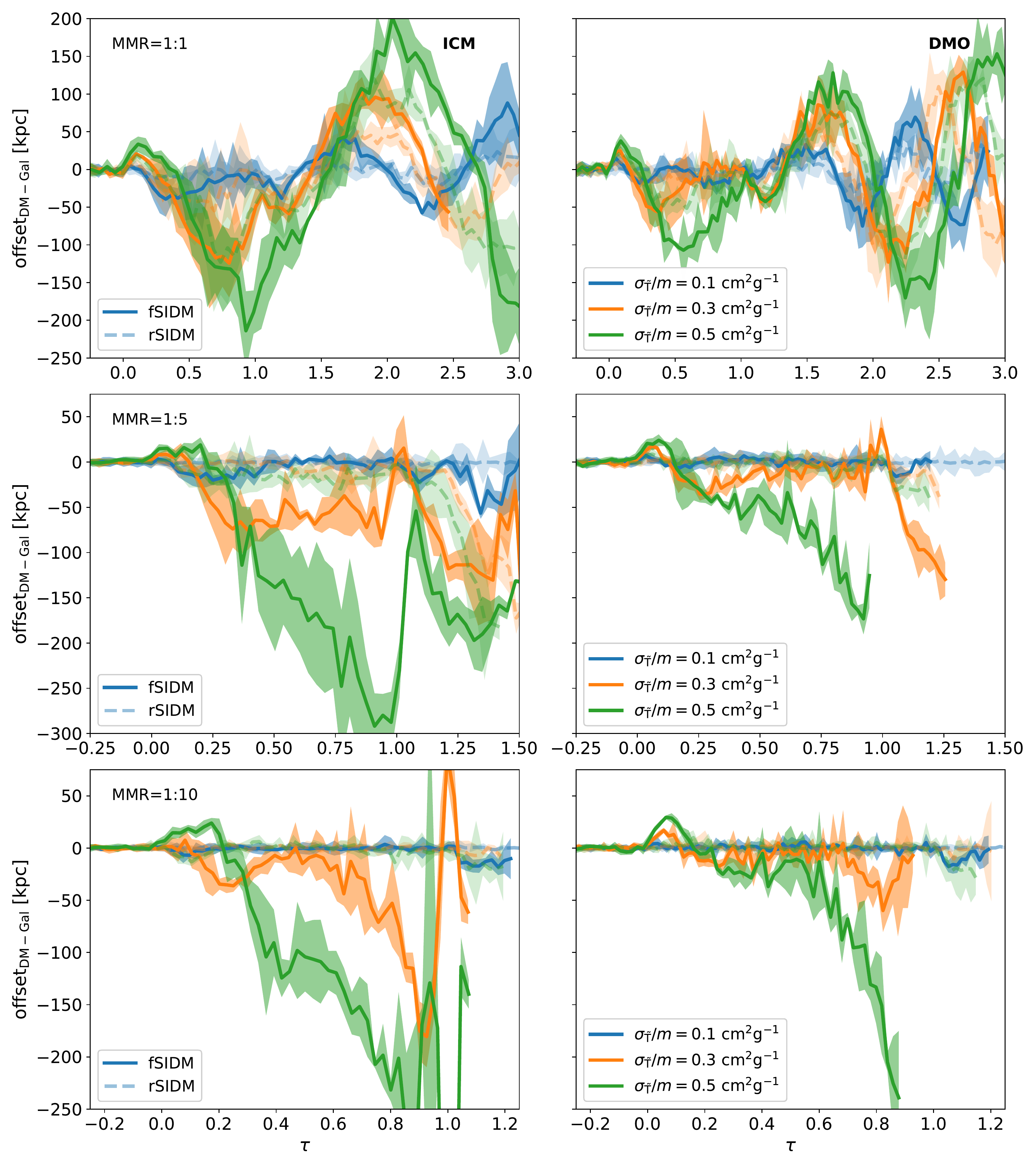}
        \caption{The offset between the DM and galactic component of the subhalo is shown for different simulations as a function of the internal merger time $\tau$ (see equation~\ref{eq:merger_time}).
        The panels on the left-hand side display the results for the simulations including the ICM and the panels on the right-hand side show the offsets of the DMO simulations \citep[previously shown in fig.~8 of][]{Fischer_2021b}.
        The top row gives the results for the equal-mass merger simulation, the middle row displays the 1:5 merger runs, and the bottom row is for the 1:10 merger.
        The colours indicate different values for the momentum transfer cross-section and the fainter dashed lines are for rSIDM and the stronger solid ones for fSIDM.
        }
        \label{fig:offsets}
\end{figure*}

\begin{figure*}
    \centering
    \includegraphics[width=\textwidth]{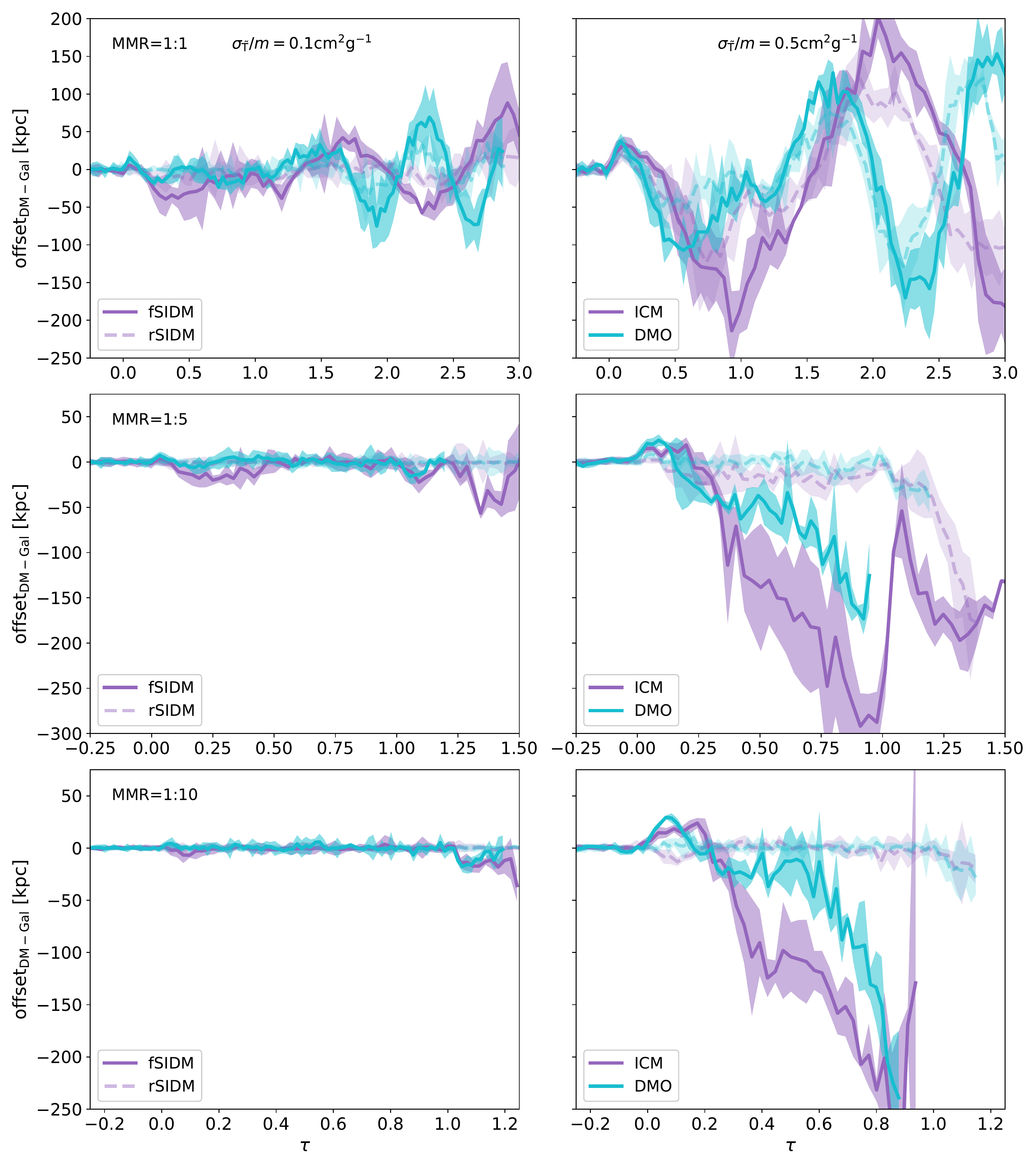}
    \caption{Similar to Fig.~\ref{fig:offsets} we show the DM--galaxy offsets as a function of the internal merger time, $\tau$.
    But here, we plot the DMO and ICM results together.
    In the left-hand column, we display the offsets for a cross-section of $\sigma_\mathrm{\Tilde{T}}/m = 0.1 \, \mathrm{cm}^2 \, \mathrm{g}^{-1}$ and on the right-hand side we show the results for $\sigma_\mathrm{\Tilde{T}}/m = 0.5 \, \mathrm{cm}^2 \, \mathrm{g}^{-1}$.
    The offsets for all merger mass ratios are given, the equal-mass merger is displayed in the top row, the 1:5 merger follows in the middle row, and in the bottom row we give the results for the 1:10 merger.
    In blue we display the offsets of the DMO runs and the simulations including the ICM are shown in purple.
    We give the results for both fSIDM and rSIDM, for the first one we use the stronger solid lines and for the latter one the fainter dashed lines.
    }
    \label{fig:offsets_comparison}
\end{figure*}

\subsection{Offsets} \label{sec:offsets}

\begin{figure*}
    \centering
    \includegraphics[width=\textwidth]{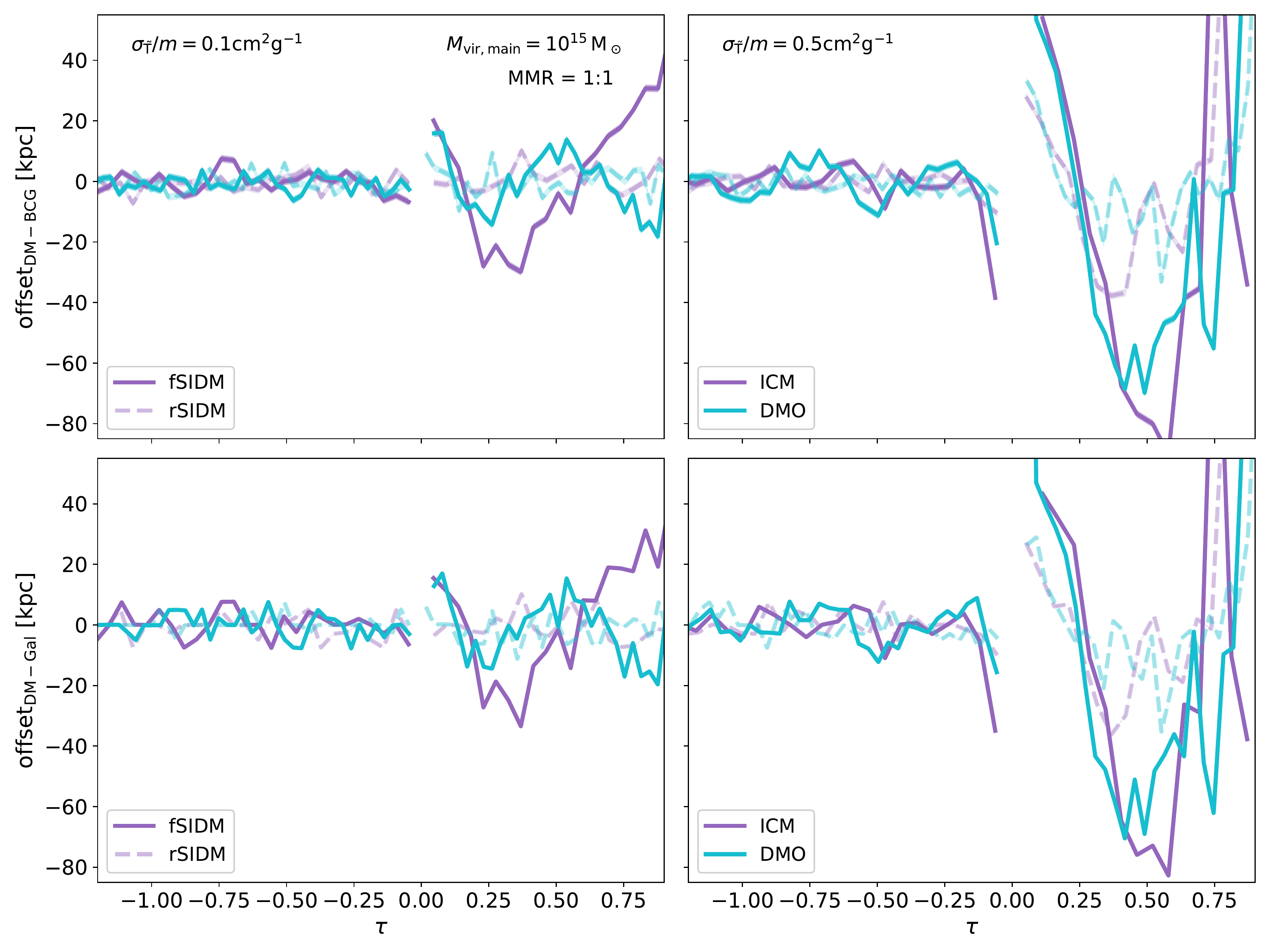}
    \caption{The DM--BCG offset (upper panels) and DM--Gal offset (lower panels) are shown as a function of the internal merger time, $\tau$ for our equal-mass mergers.
    As in Fig.~\ref{fig:offsets_comparison} we display the DMO and ICM results together.
    But this time we use the isodensity-based peaks instead of the potential-based peak finder for the DM component.
    In consequence, it is more difficult to obtain results for unequal-mass merger and we show only results for an equal mass ratio.
    For the BCG, simply the position of the particle that mimics the BCG is employed. But for the DM component and the galaxies, we are not able to determine the peaks accurately when they are close to each other. In consequence, we do not compute the offset when the DM or galaxy peaks are closer than 50 kpc.
    The left-hand panels give the results for a cross-section of $\sigma_\mathrm{\Tilde{T}}/m = 0.1 \, \mathrm{cm}^2 \, \mathrm{g}^{-1}$ and the right-hand panels show the offset for a cross-section of $\sigma_\mathrm{\Tilde{T}}/m = 0.5 \, \mathrm{cm}^2 \, \mathrm{g}^{-1}$.
    }
    \label{fig:isodens_offsets}
\end{figure*}

We compute the offsets between the DM and galactic component for our merger simulations including the ICM as previously done in \cite{Fischer_2021b}.
In addition to our ICM simulations, we also use the DMO simulations of \cite{Fischer_2021b} for comparison.

In Fig.~\ref{fig:offsets} we show the offset computed from the potential-based peaks for the subhalo.
Here, the offset is the difference between the coordinate along the merger axis, see also Eq.~\ref{eq:offset}.
On the left-hand side, we show the ICM results and on the right-hand side the DMO results.
As described in Sec.~\ref{sec:methods} the time $\tau=0$ corresponds to the first pericentre passage.
Before this time no offsets are expected and we also do not find any significant offset between the DM and galaxies.
From the first pericentre on, offsets arise due to the self-interactions which effectively decelerate the DM component.
The evolution of the offsets is roughly the same for ICM and DMO simulations.
A difficulty arises in the late stages of the merger, in particular, for the unequal-mass merger, where it becomes difficult to identify the peaks.
The subhaloes dissolve and identifying peaks from observations at these stages would be even harder.
As a consequence, the very large offsets at about the second pericentre passage and later would not be observable.
As discussed by \cite{Fischer_2021b}, the rSIDM runs show much smaller offsets than fSIDM.
For isotropic scattering, there is much less of an effective drag force that could decelerate the DM component \citep{Kahlhoefer_2014}.
However, the effective drag force does not fully vanish and a relatively small offset can occur \citep{Kim_2017b}.
When including the ICM, there is still a similarly big difference between rSIDM and fSIDM.
Often the rSIDM offsets are too small as there could be a chance to observe them.
Exceptions are only the equal-mass mergers for the larger cross-sections.

A detailed comparison between ICM and DMO simulations is displayed in Fig.~\ref{fig:offsets_comparison},
which shows the offsets from the ICM and DMO simulation for the merger with the same MMR.
For the offsets arising directly after the first pericentre passage, we do not find a significant influence of the ICM.
This agrees with the results of a previous study of a simulated Bullet Cluster analogue by \cite{Robertson_2017a}.
In contrast, the picture changes at later times, when the sign of the offset has changed.
This happens at about the first apocentre passage when the galactic peaks become closer to the barycentre than the DM peaks.
Interestingly, we find that the ICM runs show much larger offsets compared to the DMO simulations.
The difference in the trajectory of the DM and galaxies that arise at about $\tau \sim 0$, gets amplified over time and leads to large offsets at $\tau \gtrsim 0.5$.
With the ICM being present close to the centre of mass, the gravitational potential becomes deeper and the gradients steeper compared to the DMO setup.
The gravitational potential of the various components is shown in Appendix~\ref{sec:grav_pot}.
As a consequence, tidal forces are stronger, which also leads to a larger offset for the ICM runs compared to DMO.
From Appendix~\ref{sec:grav_pot}, it is visible that the ICM is more concentrated towards the barycentre than the other matter components. This leads to an increased tidal force that acts between the DM and galaxy peak compared to the DMO simulation. Thus the tidal force helps to grow the initial offset created at about the first pericentre passage by the drag force arising from the DM self-interactions.

\begin{figure*}
    \centering
    \includegraphics[width=\textwidth]{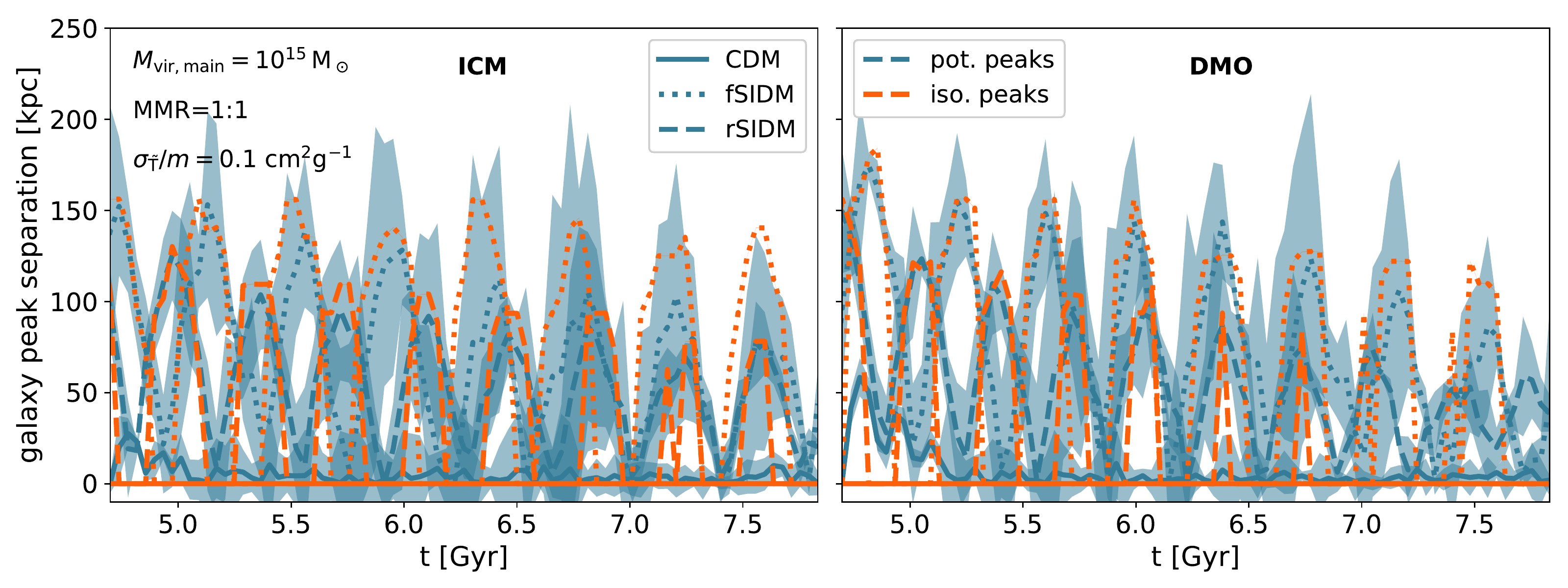}
    \caption{The separation of the galaxy peaks is shown as a function of time for the late merger stages.
    On the left-hand side, we show the results from the simulations including the ICM and on the right-hand side, the DMO runs are displayed \citep[this panel is a reprint of fig.~12 of][]{Fischer_2021b}.
    Here, we give only the results of the equal-mass mergers evolved with $\sigma_\mathrm{\Tilde{T}}/m = 0.1 \, \mathrm{cm}^2 \, \mathrm{g}^{-1}$.
    Both peak finding methods were employed and are displayed in different colours, potential-based peaks in blue and isodensity-based peaks in orange.
    }
    \label{fig:peak_separation}
\end{figure*}

Next, we investigate the offsets with the isodensity peak finder at $\tau \sim 0.5$.
In Fig.~\ref{fig:isodens_offsets} we show the offsets between DM and the BCG (upper panels) or the galaxy peaks (lower panels) for the equal-mass mergers.
In observations, it might be easier to determine the position of the BCG than trying to find a peak of a quite noisy – because of a low number – galaxy distribution.
Here, we find only a small difference between the BCG and galaxy offsets.
Only results for the equal-mass merger are shown, as the isodensity peak finding does not work well for the unequal-mass merger.
Similar to the potential-based peaks we find that the ICM does not have a significant influence on the offsets at about the first pericentre passage.
But at later stages, about the first apocentre passage, we find an enhancing effect, too.
Even for the smaller cross-section of $\sigma_\mathrm{\Tilde{T}}/m = 0.1 \, \mathrm{cm}^2 \, \mathrm{g}^{-1}$ a relatively large offset of about $\sim 30\, \mathrm{kpc}$ is visible.
For the cross-section of $\sigma_\mathrm{\Tilde{T}}/m = 0.5 \, \mathrm{cm}^2 \, \mathrm{g}^{-1}$, the offset is much larger and even the isotropic scattering leads to a significant offset.

Current constraints from density cores of relaxed galaxy clusters at a confidence level of 95\% are, $\sigma_\mathrm{\Tilde{T}}/m = 0.175 \, \mathrm{cm}^2 \, \mathrm{g}^{-1}$ from \cite{Sagunski_2021} and $\sigma_\mathrm{\Tilde{T}}/m = 0.065 \, \mathrm{cm}^2 \, \mathrm{g}^{-1}$ from \cite{Andrade_2021}.\footnote{Here, we assumed that we can transfer limits for isotropic scattering to fSIDM by using the momentum transfer cross-section, for detailed discussion on this approach see section~3.10 of \cite{Fischer_2022}}
Given these limits, it does not seem to be excluded that a merger could have a significant DM--galaxy offset.

\subsection{Core sloshing} \label{sec:core_sloshing}

We also study very late merger phases when the DM components have coalesced.
It has been found by \cite{Kim_2017b} that the BCGs at late stages oscillate and their amplitude is only very slowly decreasing.
This is mainly caused by a reduced dynamical friction force as a consequence of a lower matter density \citep{Fischer_2021b}.
The same is true when considering the peaks of the smooth galactic component.
In Fig.~\ref{fig:peak_separation} we show the separation between the galaxy peaks at late times of the equal-mass merger for both runs ICM and DMO.
The distance between the two peaks of the galaxy distribution is displayed for both peak finding methods, the potential-based and the isodensity-based peaks.
At this evolution stage, no separate galaxy peaks exist for the CDM runs.
But both SIDM runs show core sloshing, independent of the angular dependence of the differential cross-section.
Furthermore, we notice that for fSIDM, the amplitude is a bit larger than for rSIDM but we do not find a significant influence of the ICM on the core sloshing.
Previously, it was not clear whether it still would be present when including the ICM.
However, we have to note that these simulations are idealized and a more realistic treatment of the substructure within the galaxy cluster could affect the core sloshing.
We also do not show the results from the unequal-mass mergers as they are not relaxed in our simulations and we would not be able to identify the galaxy peaks at that late stages.
\cite{Harvey_2019} studied BCG oscillations in BAHAMAS-SIDM \citep[a cosmological hydrodynamic simulation,][]{Robertson_2019} and did not find a significant deviation between CDM and observations.
But the BCG oscillations allowed them to place tight constraints on the self-interaction cross-section ($\sigma/m < 0.39 \, \mathrm{cm}^2 \, \mathrm{g}^{-1}$ at 95\% confidence level for the total cross-section) and might be of interest for future studies.

\subsection{Morphology} \label{sec:morphology}

\begin{figure*}
    \centering
    \includegraphics[width=\textwidth]{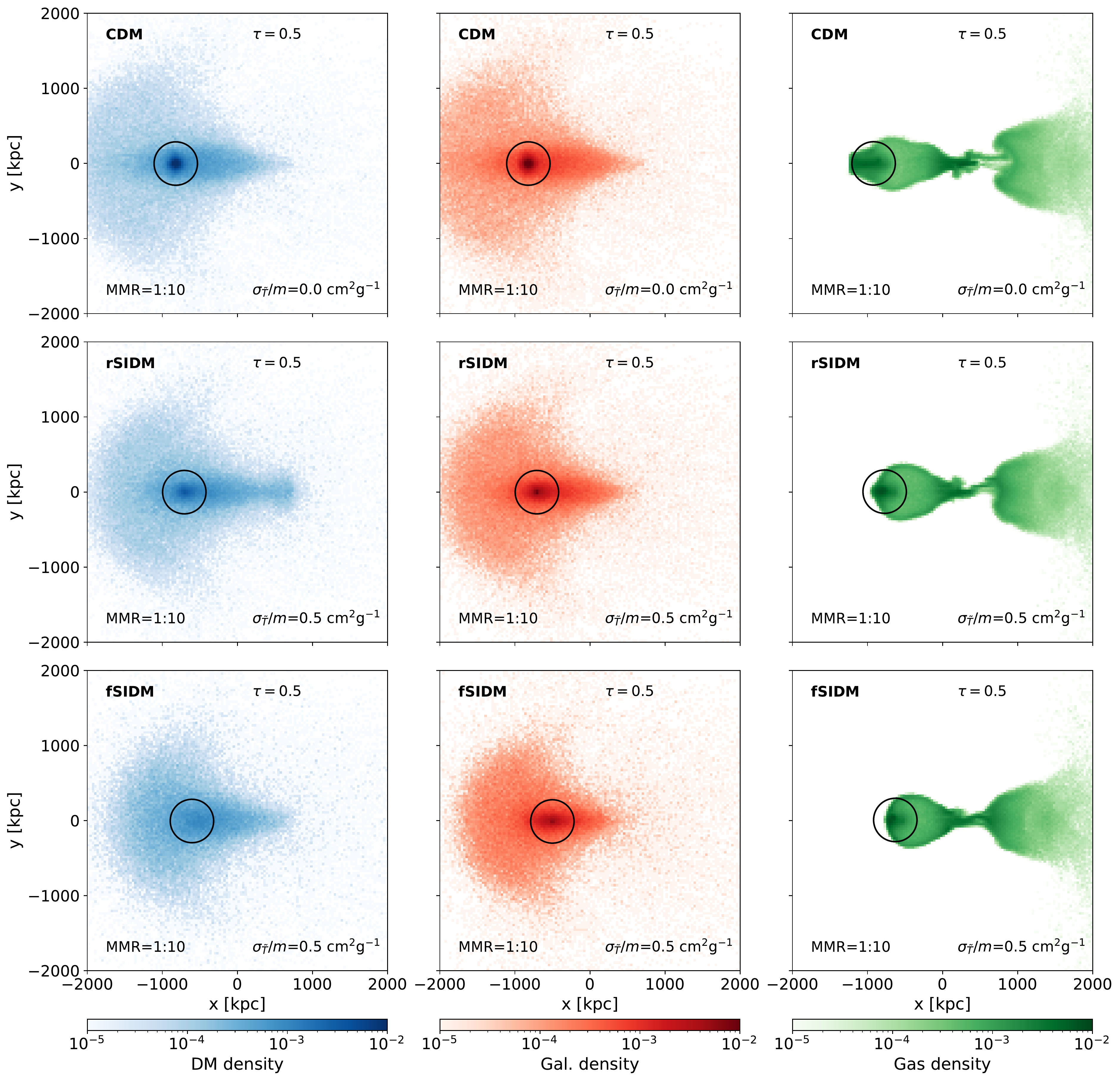}
    \caption{We show the morphology of the subhalo of our 1:10 merger at $\tau=0.5$, roughly speaking the first apocentre passage.
    The physical density in the merger plane of the various components of the subhalo only is displayed.
    Note, we normalized it using the mass of each matter component.
    The DM is shown in blue, the galaxies in red, and the ICM in green.
    All densities have been computed by using only the particles that initially belong to the smaller halo.
    The peaks are marked by the black circles, which have a radius of $2 r_s$.
    The top row gives the results for CDM, the middle row for rSIDM, and the bottom row for fSIDM.}
    \label{fig:morphology}
\end{figure*}

\begin{figure*}
    \centering
    \includegraphics[width=0.95\textwidth]{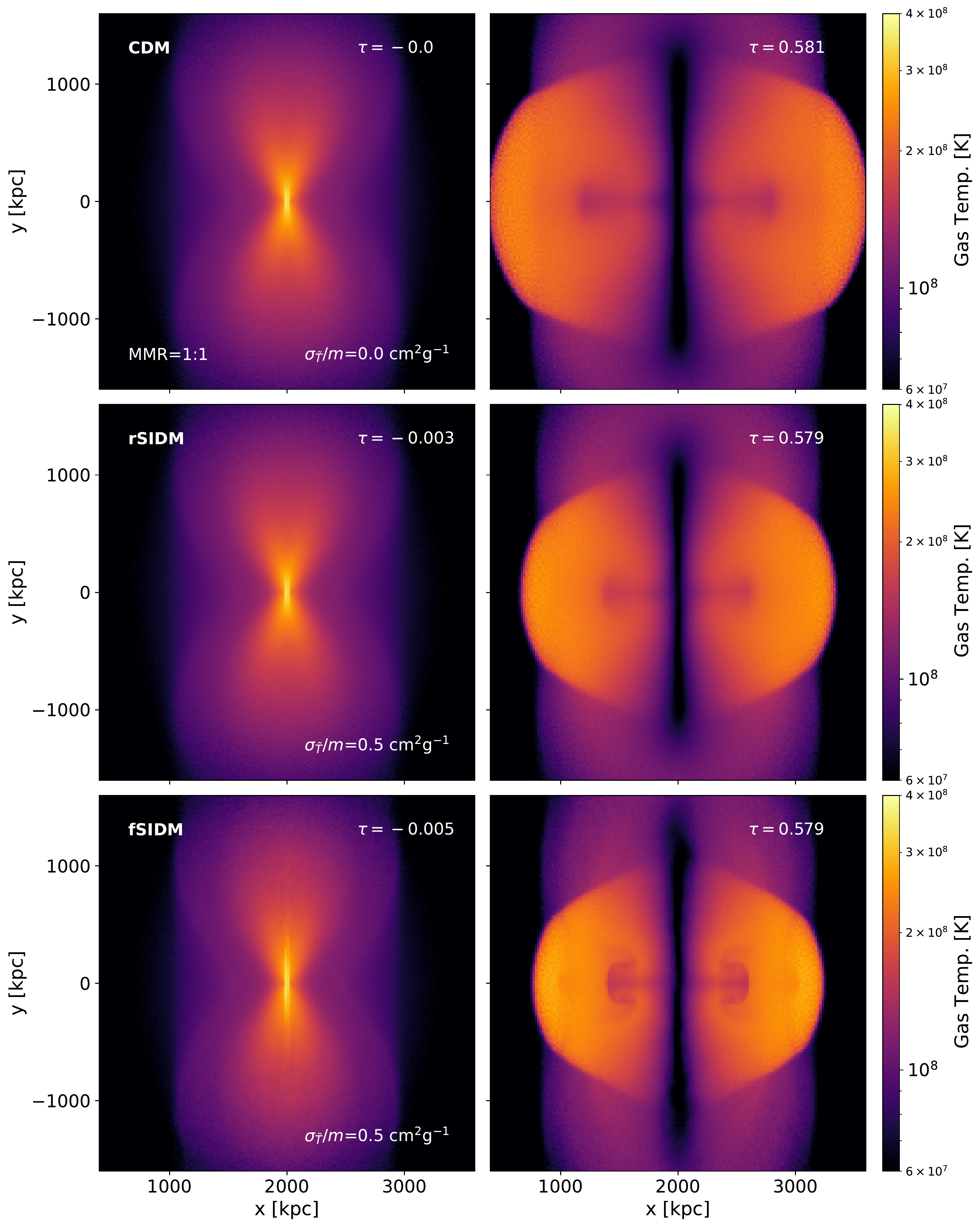}
    \caption{
    A mass-weighted temperature map projected perpendicular to the merger axis is shown.
    The results are for the equal-mass merger evolved with CDM (top row), rSIDM (middle row), and fSIDM (bottom row).
    The left-hand side displays the temperature at the first pericentre passage ($\tau \approx 0$).
    A stage shortly after the first apocentre passage ($\tau \approx 0.58$) is shown on the right-hand side.
    }
    \label{fig:temp}
\end{figure*}

So far we have described the matter distribution by approximating it with a single point, the peak position.
Higher-order moments of the distribution may contain valuable information, too.
For example, tidal features could also turn out to help distinguish DM models.

In Fig.~\ref{fig:morphology} we show the physical density in the merger plane using only particles that initially belong to the subhalo.
It becomes clear that the CDM halo is more concentrated than the SIDM counterparts.
For fSIDM, the subhalo dissolves the fastest and hence has the lowest central density.
This is not only true for DM but also for the galactic component.
Moreover, the distributions do not only differ in concentration.
But we can also see that for rSIDM more DM particles are captured by the main halo than in the other DM models.
Furthermore, clear morphological differences between the DM models are visible in the gas density.
However, in observations, such features might be more difficult to observe as they could be masked by the main halo.
Moreover, in projection, the morphology may look somewhat different than the physical density in the merger plane.

In addition to the density, we also study the temperature of the ICM in projection and for the equal-mass merger.
In Fig.~\ref{fig:temp} we show the mass-weighted temperature along a line-of-sight that is perpendicular to the merger axis.
In order to compute the temperature from the internal energy that \textsc{gadget-3} provides for the SPH-particles, we assume for the ICM an adiabatic index of $\gamma = 5/3$ and a mean molecular weight of $\mu \approx 0.59$.
The merger is shown at different times, starting with the first pericentre passage, $\tau = 0.0$, on the left-hand side.
A stage shortly after the first apocentre passage, $\tau=0.58$, is displayed on the right-hand side.\footnote{We select those times as we have snapshots of all three models that correspond closely to those merger times. However, this implies that they are not exactly for the same merger time but with a deviation of not more than $\Delta \tau = 0.005$. In contrast to Fig.~\ref{fig:morphology}, we want to avoid interpolation as the shock fronts propagate quickly.}

At about the first pericentre passage the DM models look fairly similar but differences in the temperature for the fSIDM run compared to the others are visible.
As a result of the collision, the gas is heated, a shock wave arises and travels outwards.
At later times, i.e.\ larger values for $\tau$, differences between the DM models become larger.
Especially, at stages as late as the first apocentre passage.
In Fig.~\ref{fig:temp} we can see significant differences between the positions of the shock fronts.
For CDM the shock front has travelled the furthest outside, less for rSIDM, and in the fSIDM run the shock front has travelled the least far.
It is worth mentioning that this is in terms of the internal merger time $\tau$, i.e.\ comparing the same evolutionary stages based on the BCG positions.
The picture changes when comparing the runs at the same physical time instead of using $\tau$.
In section~\ref{sec:shock} we investigate the distance of the shock front in more detail.
In addition, the fSIDM runs show a more detailed jellyfish or mushroom-like structure of colder gas far behind the shock front.

These results seem promising that the ICM could help to discriminate between DM candidates.
In particular, the position of the shock front is propitious, which we will investigate next.
But a more detailed study would be needed to see how well the morphological features in density and temperature can be used to distinguish between DM models and constrain their parameters.

\subsection{Position of the shock front} \label{sec:shock}

\begin{figure}
    \centering
    \includegraphics[width=\columnwidth]{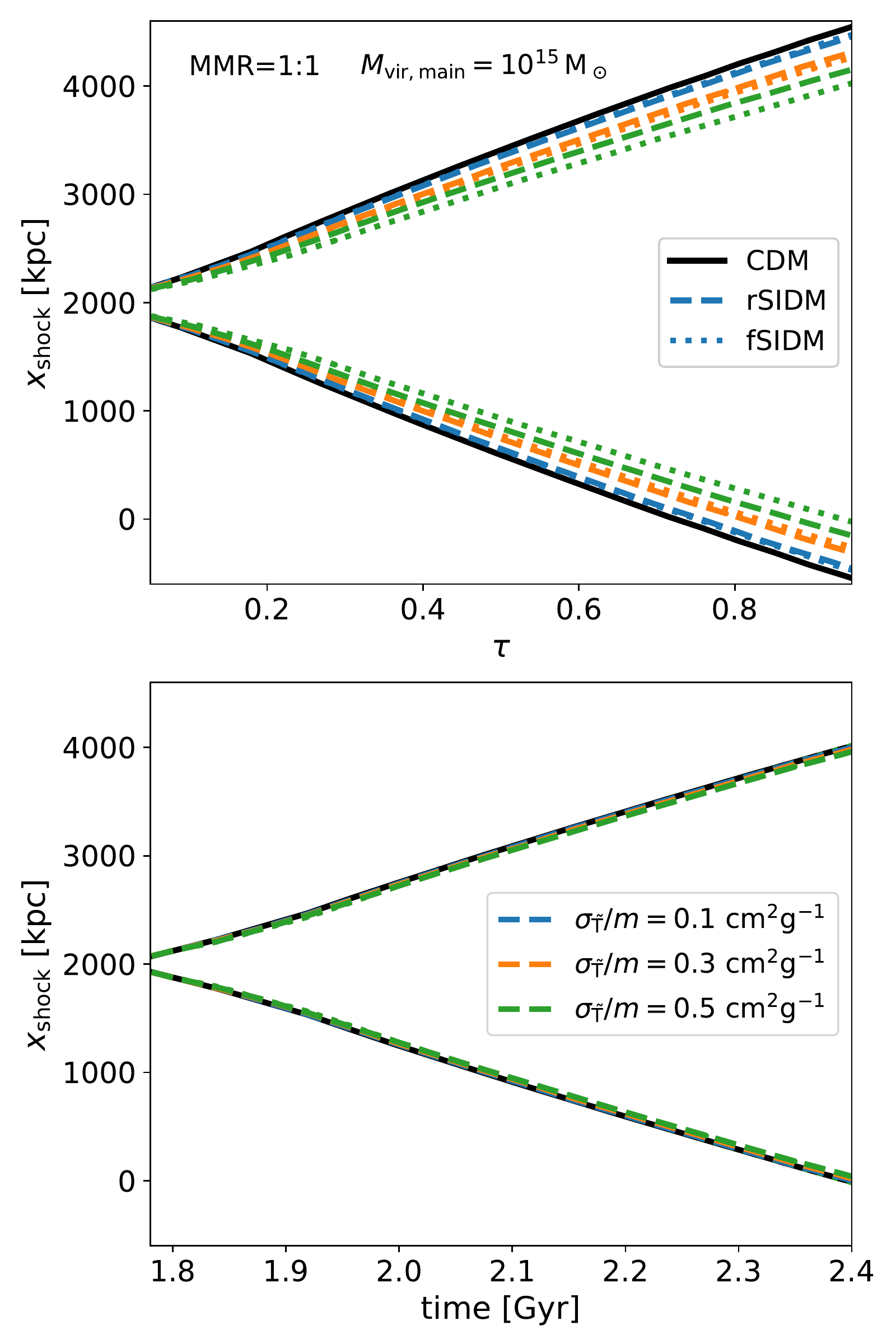}
    \caption{We show the position of the shock front as a function of the internal merger time $\tau$ (upper panel) and the physical time (lower panel) for the equal-mass merger. The evolution of the shock front for the different DM models is displayed in different colours and line types as indicated in the legend.}
    \label{fig:shock_pos}
\end{figure}

\begin{figure}
    \centering
    \includegraphics[width=\columnwidth]{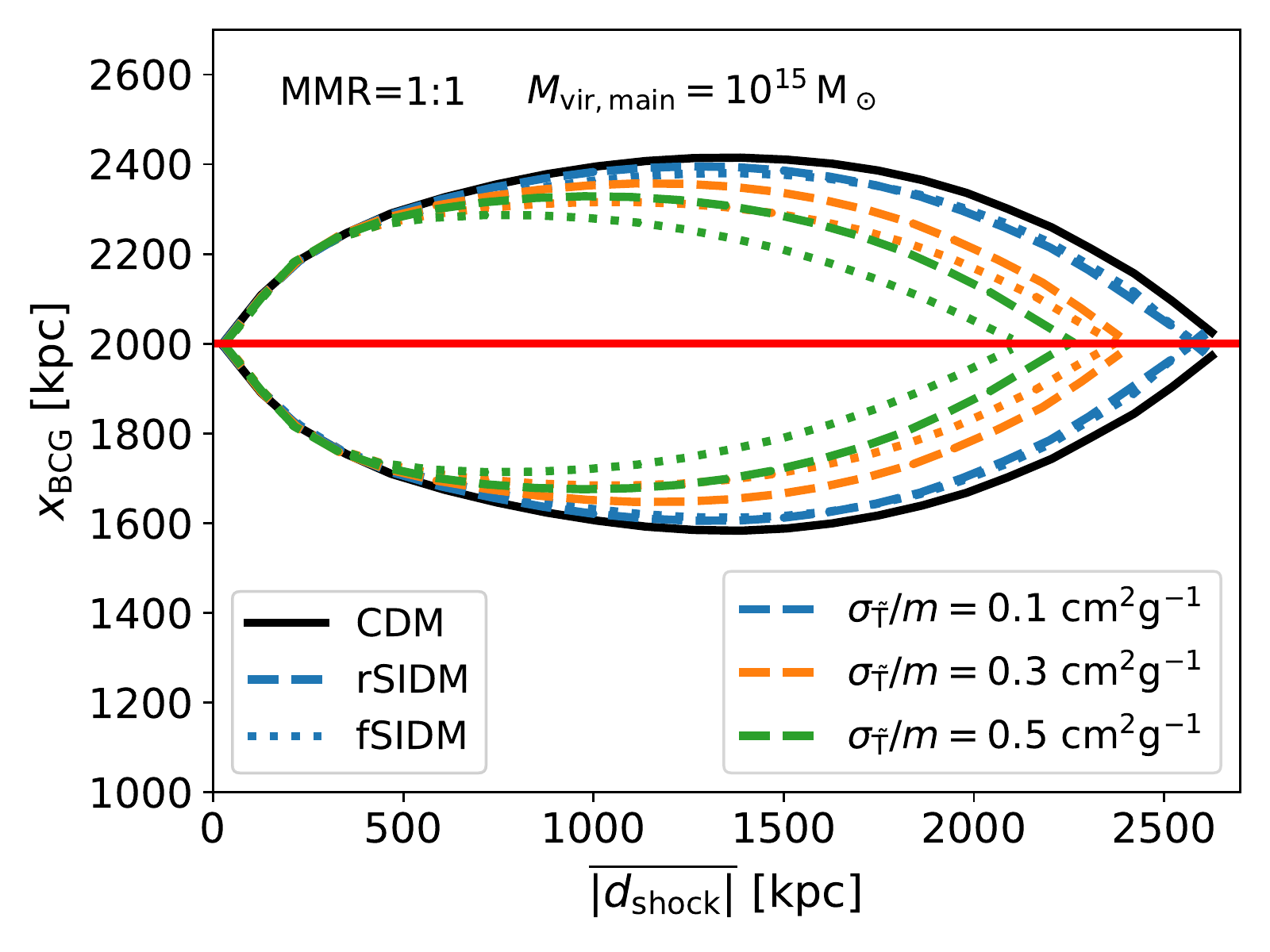}
    \caption{The position of the BCGs for the equal-mass merger is shown as a function of the distance that the shock fronts have travelled. Here, we give the results for the period between the first and second pericentre passage. The distance, $\overline{|d_\mathrm{shock}|}$ is the average of both shocks. We display the results for different DM models as specified in the legend. The centre of mass of the merger is indicated by the red line.}
    \label{fig:bcg_shock}
\end{figure}

In Fig.~\ref{fig:temp} we found the position of the shock front to be of interest.
Here we measure the position of the shock front along the merger axis.
We do this in three dimensions based on temperature gradients and show our findings in Fig.\ref{fig:shock_pos}.
The upper panel gives the shock position in terms of the internal merger time $\tau$.
Here we can see that for larger cross-sections the shock front is located further inwards.
The lower panel displays the shock position as a function of the physical time.
Only very slight differences between the DM models are visible.
Hence, the position of the shock front is almost insensitive to the DM model, at least for the models we tested here.
Close to the first pericentre passage the difference in the gas distribution and the gravitational potential between the models is small and does not affect the shocks largely.
It is also important, that the infall velocity of the haloes is independent of the DM model in our setup and thus does not play a role in the morphological differences we find here.
At later stages, the shock fronts have travelled far outside beyond the scales that are significantly altered by DM self-interactions. In other words, the fast propagation of the shock fronts lets them leave the inner region of the clusters before they are largely affected by the self-interactions. This makes them almost independent of DM physics.
However, we can only make this conclusion for the merger set-up we have studied here and the 1:5 merger.
For the 1:10 merger, we start to see an influence of the DM physics.
The differences in the shock position shown in the upper panel are primarily due to the shorter merger times caused by DM self-interactions.
That self-interactions do affect the merger time, is for example visible in Fig.~\ref{fig:peakpos_equal}.

The positions of the shock fronts are mainly independent of DM physics.
Combined with other matter components, this could make them a promising tracer for the nature of DM.
To show this in detail we plot the positions of the BCGs between the first and second pericentre passage as a function of the distance of the shock front.
For the later we introduce,
\begin{equation}
    \overline{|d_\mathrm{shock}|} \equiv \frac{|x_{\mathrm{shock},1} - x_{\mathrm{shock},2}|}{2} \,.
\end{equation}
Where $x_{\mathrm{shock},i}$ denotes the position of the individual shock fronts along the merger axis.
The results are displayed in Fig.~\ref{fig:bcg_shock} and can be seen in analogy to Fig.~\ref{fig:peakpos_equal} but focused on the phase between the first and second pericentre passage with the BCGs only but therefore multiple DM models.
At the phase, shortly after the first pericentre passage all DM models behave fairly similarly.
But at later stages, the differences grow.
The stronger the self-interactions the shorter the merger time and the smaller the merger amplitude are for the BCGs.
Given the position of the shock fronts and the BCGs, this could allow to constrain the strength of DM self-interactions.
However, how applicable this is for real galaxy clusters and their observations still needs to be investigated.

\section{Discussion} \label{sec:discussion}

In this section, we discuss technical aspects regarding the modelling and analysis of our galaxy cluster simulations and elaborate on the physical implications and limitations of our study.

\subsection{Technical aspects}

We employed two peak-finding methods, one based on the gravitational potential and the other one based on isodensity contours.
The first one is more robust in identifying the peaks and thus allows us to study them at later times.
Hence, it is preferable to study qualitative differences between DM models.
Nevertheless, the peak finding for unequal-mass mergers is in general difficult as the subhalo dissolves quickly.
The method based on isodensity contours is less robust but closer to observations and thus gives a better idea of what could be observable.

But we have to note, in real observations further difficulties arise as real galaxy clusters contain much fewer galaxies compared to the number of particles we used to model a smooth galactic component.
The number of observed galaxies is only about $\sim$ 100 -- 1000 for a galaxy cluster.

Although peak finding at later merger stages faces huge challenges, one may observe core-sloshing.
The BCGs of the progenitor clusters oscillate and their offset from the centre can be observable.
For rSIDM, this has been studied by \cite{Harvey_2019}.
They find much smaller oscillation amplitudes ($\lesssim 40\,\mathrm{kpc}$) than we do. In contrast to our work, they study galaxy clusters from a cosmological SIDM simulation \citep{Robertson_2019}. Their clusters are less massive ($\approx 2.5 \times 10^{14} \, \mathrm{M_\odot}$) than ours and the merger mass ratio and the in-fall orbits are set by cosmological structure formation which makes events similar to the setup we have studied fairly uncommon. Instead, our work gives an idea of what the maximum oscillation amplitude in a tuned system could be but does not inform us about the distribution of oscillation amplitudes that we would expect from realistic systems. A merging system such as the one we have studied is rare in both cosmological simulations and observations. Nevertheless, observational studies of the BCG oscillations increase in sample size \citep{Cross_2023} and may grow further thanks to new observations such as high-resolution X-ray images or cluster lensing surveys. In particular \textsc{Euclid} \citep{Laureijs_2011, Euclid_Collaboration_2022} is expected to image $\sim 10^3$--$10^5$ clusters.

Finally, we studied morphological differences between the DM models.
The differences in the distribution of the galaxies might be hard to observe as the number of galaxies in clusters may be insufficient to see the effect. However, the differences in the ICM might be more promising to constrain DM models.
Especially when combining the distance of the shock fronts with the positions of the BCGs.
The position of the BCG can be determined quite well, whereas the position of the shock front is more difficult to detect. To date, there are more than 20 galaxy clusters with confirmed shock fronts known \citep[e.g.][]{ZuHone_2022}. It is easier to determine their position if they are close to the cluster centre and thus X-ray bright. However, we are more interested in shock fronts that are located in cluster outskirts. Given that they host radio relics their position can be precisely determined \citep[e.g.][]{Bruggen2020, Rajpurohit2021}.
Before turning this measure into a probe of potential DM physics one would need to understand it in greater detail from a theoretical perspective. In contrast to the DM--galaxy offsets, it is less clear what the expectation for CDM is. The distance between the BCG and the shock front may depend on the exact merger configuration, which can be subject to substantial uncertainties for comparison based on individual objects. To take this forward, it would be reasonable to understand how uncertainties in the merger configuration translate to different BCG--shock front distances. This would allow us to understand how sensitive this probe can be for new DM physics and how promising a comparison of individual objects or a statistical approach would be.

\subsection{Physical considerations}

In this section, we revisit physical assumptions made throughout this paper and discuss their validity. We want to point out that we made those assumptions in line with the paper's aim to provide a better qualitative understanding of the role of the ICM in mergers of galaxy clusters with SIDM and not a quantitative comparison with observations.
We will discuss how breaking assumptions could lead to different results.

Throughout this paper, we have implicitly assumed that it is reasonable to compare the DMO and ICM runs with each other.
The galaxy clusters including the ICM have the same mass as their DMO counterparts.
Only fractions of the different mass components are different.
In addition, the ICM follows a different density profile than the other components, which leads to a slightly reduced central density for the haloes.
This could affect the relative velocity of the central regions of the merging galaxy clusters at the first apocentre passage.
A slower relative velocity would lead to smaller offsets \citep{Kim_2017b}.
In Fig.~\ref{fig:offsets_comparison} we found that the size of the offsets shortly after the first pericentre passage have a very similar size for the DMO and ICM runs.
This makes the simulations quantitatively comparable for the early merger phase. It also becomes clear that the later offsets are qualitatively different when including the ICM.

Although we included the ICM component in our simulations, for a detailed comparison with observations more sophisticated simulations are required, because several physical aspects have been ignored.
For the sake of a qualitative study, it makes sense not to include all physical effects at once to see qualitative differences in plain sight. However, follow-up studies may include some aspects we have neglected here.
These are the substructure within the cluster, the cooling of the ICM through radiative losses and the influence of active galactic nuclei, to name only the most important ones.
For example, the fact that individual galaxies can have a DM halo has been studied by \cite{Kummer_2018}.
Furthermore, a galaxy is an extended object and approximating the BCG as a point-like mass might be quite inaccurate.

In addition to this, comes the dependence on the exact merger configuration.
Various physical processes act on different timescales which depend on the chosen setup.
In consequence, there is some variation in the merger phenomenology, which we did not cover here.
Given, a suitably large number of mergers, we could determine how frequent observable offsets would be given a specific DM model.
In fact, many mergers would have a non-vanishing impact parameter, which we expect to make large offsets rare.
This is because the impact of the self-interactions on the merger evolution depends on the DM density too.
If the haloes do not pass through the centre of each other, i.e.\ in the case of a significant impact parameter, lower densities are involved and thus the effect of SIDM should be weaker, implying smaller offsets.

Aside from the astrophysical idealizations, the particle physics assumptions play a major role here.
It is worth pointing out that we used the momentum transfer cross-section to match rSIDM and fSIDM.
Matching the cross-sections is in general problematic as no perfect matching procedure exists.
For our purpose, the momentum transfer cross-section should be reasonable, as the matched differential cross-sections result in roughly similarly strong effects on the DM distribution.
Nevertheless, we may have to revisit this for a quantitative study and may consider a different approach.
This could be done similarly to the one described in Sec.~3.9 of \cite{Fischer_2022} by using a different measure such as the cores size of relaxed DM haloes to match DM models.

In this study, we have only considered velocity-independent elastic scattering.
However, DM candidates with a velocity dependence are well motivated by particle- and astrophysics.
Especially light-mediator models, which fall into the fSIDM class, would interact velocity dependent \citep[e.g.][]{Buckley_2010, Loeb_2011, Bringmann_2017}. 
Moreover, cross-sections that are decreasing with velocity appear to be in better agreement with astronomical observations than velocity-independent scattering \citep[e.g.][]{Kaplinghat_2016, Correa_2021, Gilman_2021, Sagunski_2021}.
Finally, there are further model variations that could be studied in the context of mergers, for example, dissipative self-interactions.

\section{Conclusion} \label{sec:conclusion}

We studied equal- and unequal-mass mergers of galaxy cluster head-on collisions with rSIDM and fSIDM in order to understand the qualitative difference that the ICM makes in those two scenarios.

To this end we ran idealized SPH + $N$-body simulations with \textsc{gadget-3}.
We identified the peaks of the matter components (DM, galaxies, ICM) using different peak-finding methods and studied the evolution of the peak positions as well as offsets between DM and galaxies.
Furthermore, we investigated the morphology of the mergers in terms of density and temperature distribution as well as the position of the shock front.
Our most important results can be summarized as follows:

\begin{itemize}
    \item The ICM affects the peak positions of the matter components significantly. It shortens the merger time and the merger amplitude.
    \item The ICM increases the offsets significantly compared to DMO simulations after the first pericentre passage.
    \item A cross-section small enough not to be ruled out by current constraints could produce an observable DM--galaxy offset.
    \item BCG/galaxy oscillations in the merger remnant (late evolutionary stages) are hardly impacted by the presence of the ICM.
    \item The shock fronts in the ICM are mostly insensitive to the DM models we studied. Combining this with a DM model-sensitive measure, such as the position of the BCGs, could help distinguish DM models at merger stages well after the first pericentre passage.
\end{itemize}

Perhaps the most important results are the significant offsets for small cross-sections and the mostly DM model-independent propagation of the shock front.
Even given the current upper limits on the self-interaction cross-section, there could be a chance that potential offsets are observable with forthcoming telescopes.
But further, more detailed studies, such as cosmological simulations, are needed.
In summary, mergers of galaxy clusters seem to be an interesting test bed to study the nature of DM.
They could allow placing independent constraints on frequent self-interactions with the chance to be similarly tight as current upper limits on the cross-section from density cores.
In case, a significant offset is observed, this allows to place a lower limit on the fSIDM cross-section and would make frequent self-interactions a very promising DM candidate.

It is worth studying mergers near the first apocentre passage when offsets are larger and more persistent.
So far, SIDM simulations of cluster mergers including baryons have only studied Bullet Cluster analogous shortly after the first pericentre passage \citep{Robertson_2017a}.
The result that the ICM can increase offsets motivates further work.
From the observational side, this would include searching for promising merging clusters and striving for more accurate measures of DM--galaxy offset.
In order to derive theoretical predictions for potential offsets, it would be reasonable to run cosmological simulations and understand how rare large offsets are.

Another avenue worth looking into is using the morphology of the ICM to study DM properties.
In particular, the position of the shock fronts might be interesting, not because they are strongly affected by DM physics but because they are very weakly affected only.
Combining them with a measure strongly influenced by DM physics could allow constraining DM models. Using the BCGs would not even involve gravitational lensing.
Yet it is to be explored how well such a combination can tell DM models apart.
But we found, similar to DM--galaxy offsets, later merger stages are more sensitive to DM self-interactions.

Lastly, we want to mention the oscillations of the BCGs/galaxies in the merger remnant. They may provide a promising probe for SIDM. Improving on the work by \cite{Kim_2017b} we found that larger oscillation amplitudes can be still present when including the ICM. For more common merger configurations the oscillation amplitude would be much smaller as found in cosmological SIDM simulations \citep{Harvey_2019}. In particular, growing sample sizes from both observations and simulations may allow to better constraint the number of clusters with large BCG offsets and to deduce bounds on the DM self-interaction cross-section.

Eventually, future studies based on a detailed comparison of simulations and observations may provide further insights into the nature of DM.

\section*{Acknowledgements}
We would like to express our appreciation and respect to Nils-Henrik Durke, who unfortunately passed away in February 2022. He was a master's student at the University of Hamburg and started this project as his master’s thesis. We would have loved to see him finish the simulations, analyse the data, and publish the results. Unfortunately, he did not have this opportunity. We remember a kind person and a gifted student.

We want to thank the anonymous referee for helpful comments that improved the paper.
This work is funded by the Deutsche Forschungsgemeinschaft (DFG, German Research Foundation) under Germany's Excellence Strategy – EXC 2121 ``Quantum Universe'' –  390833306 and Germany’s Excellence Strategy – EXC-2094 ``Origins'' – 390783311.
KD acknowledges support by the COMPLEX project from the European Research Council (ERC) under the European Union’s Horizon 2020 - Research and Innovation Framework Programme grant agreement ERC-2019-AdG 882679.
The simulations have been carried out on the computing facility Hummel (HPC-Cluster 2015) of the University of Hamburg.

Software:
NumPy \citep{NumPy},
Matplotlib \citep{Matplotlib}

\section*{Data Availability}

The data underlying this paper will be shared on reasonable request to the corresponding author.



\bibliographystyle{mnras}
\bibliography{references} 


\section*{Supporting Information}
Supplementary data are available at \href{https://academic.oup.com/mnras/article-lookup/doi/10.1093/mnras/stad1786#supplementary-data}{\textit{MNRAS}} online.
\newline

\noindent Please note: Oxford University Press is not responsible for the content or functionality of any supporting materials supplied by the authors.
Any queries (other than missing material) should be directed to the corresponding author for the paper.


\appendix

\section{Stability of initial conditions}
\begin{figure*}
    \centering
    \includegraphics[width=\textwidth]{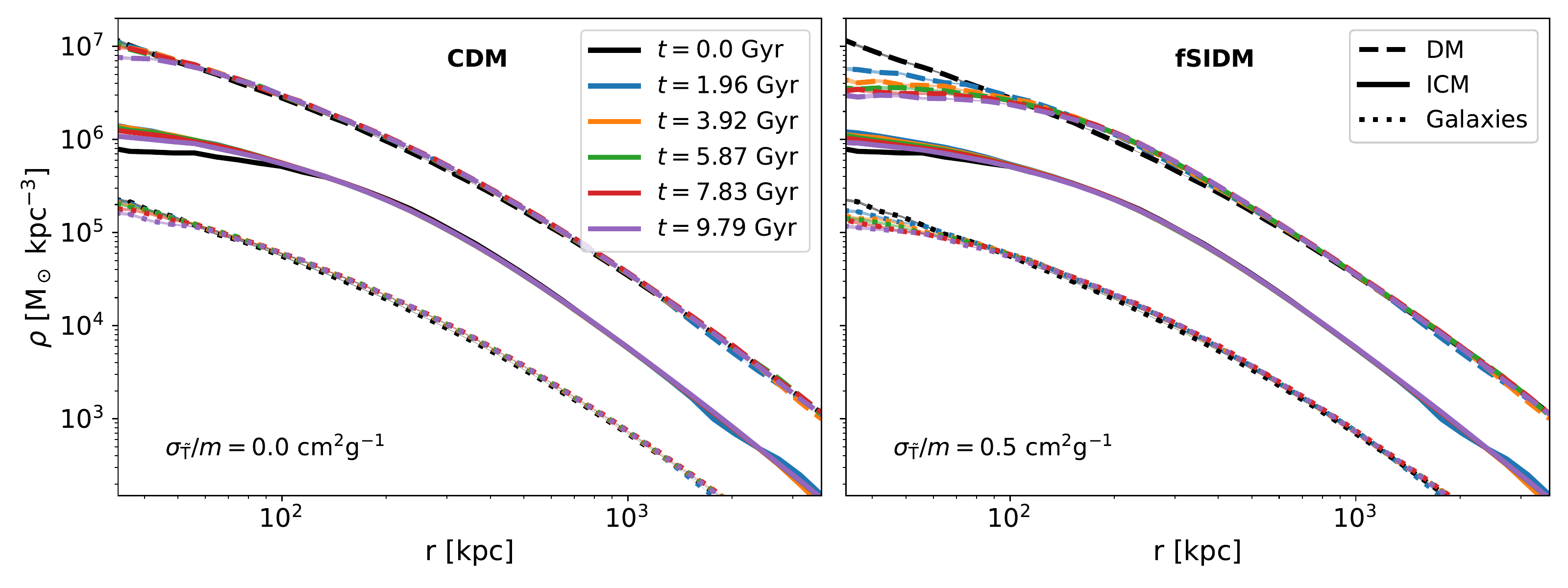}
    \caption{We show the density as a function of radius for the main halo ($M_\mathrm{vir} = 10^{15} \, \mathrm{M_\odot}$) of our merger simulations when evolved in isolation. The left-hand panel gives the run for CDM and the right-hand panel gives the run for fSIDM with a cross-section of $\sigma_\mathrm{\Tilde{T}}/m = 0.5 \, \mathrm{cm}^2 \, \mathrm{g}^{-1}$.}
    \label{fig:halo_profile}
\end{figure*}

In this appendix, we demonstrate that the individual NFW haloes of our ICs are stable.
We performed simulations of the main halo ($M_\mathrm{vir} = 10^{15} \, \mathrm{M_\odot}$) including the ICM.
In Fig.~\ref{fig:halo_profile}, we show the density profile of the halo at several times for CDM (left-hand panel) and fSIDM with a cross-section of $\sigma_\mathrm{\Tilde{T}}/m = 0.5 \, \mathrm{cm}^2 \, \mathrm{g}^{-1}$ (right-hand panel).
For the ICM, the inner density quickly increases and stays roughly constant over time. This is caused by the ICs not being in perfect equilibrium.
We can see that for CDM only a small numerical core forms at late times and besides this the ICs are stable. Hence, they behave as supposed and are suited for our study.
The fSIDM run shows a larger density core as the self-interactions transfer heat to the central region of the halo over time.
This is in line with numerous previous studies of SIDM \citep[e.g.][]{Burkert_2000, Dave_2001, Fischer_2021a}.

\section{Gravitational potential} \label{sec:grav_pot}
\begin{figure*}
    \centering
    \includegraphics[width=\textwidth]{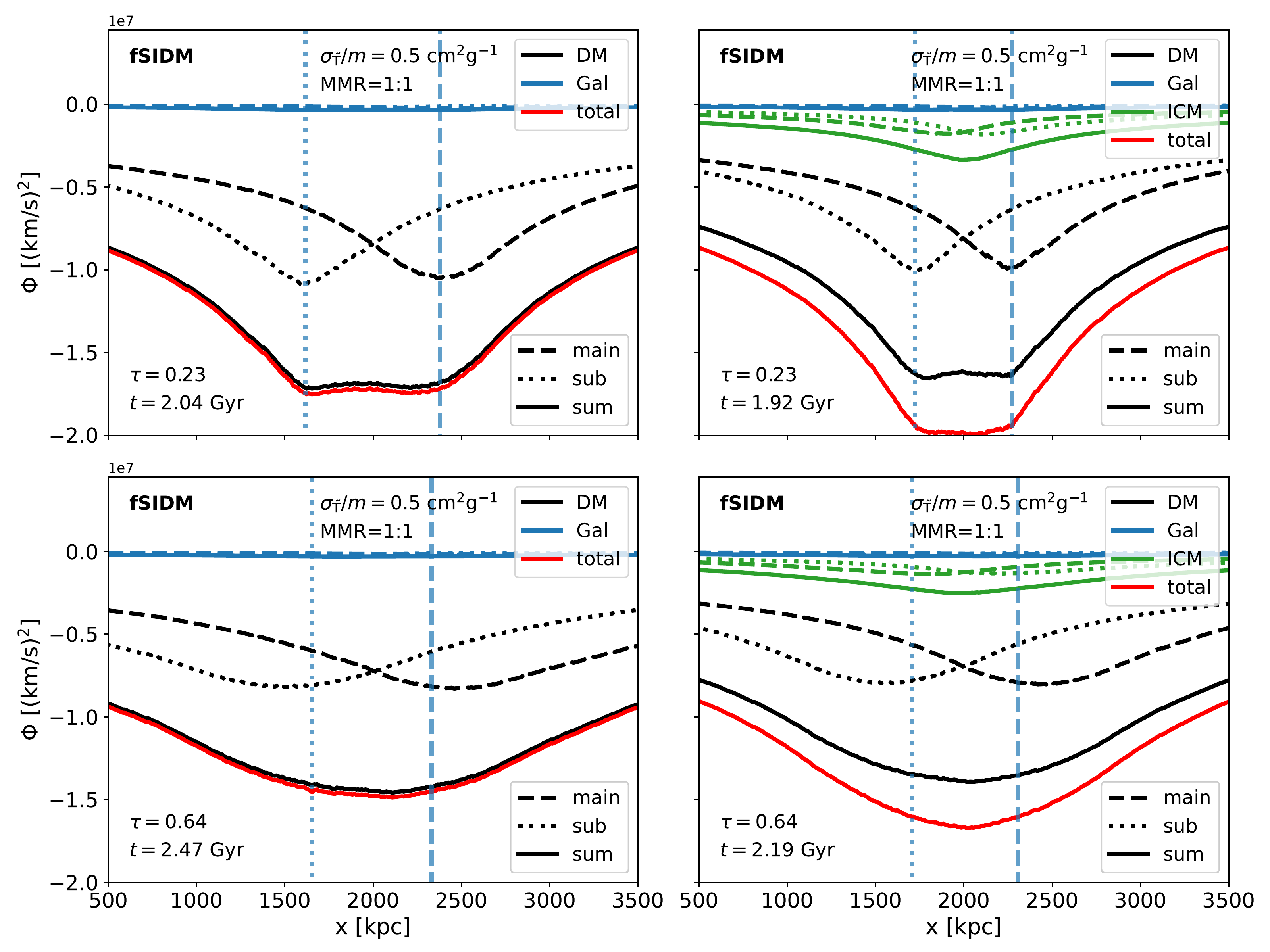}
    \caption{The gravitational potential along the merger axis is shown for the different components of our merger simulations. We display the results of the equal-mass merger evolved with frequent self-interactions using a cross-section of $\sigma_\mathrm{\Tilde{T}}/m = 0.5 \, \mathrm{cm}^2 \, \mathrm{g}^{-1}$. We indicate the peak position of the galaxies with the vertical lines. The left-hand panel displays the result of the DMO simulation of \protect\cite{Fischer_2021b} and the right-hand panel gives the new simulation including the ICM. The top row gives the results for $\tau = 0.23$ and the bottom row for $\tau=0.64$.
    In the supplementary material, we provide the time evolution as a video.
    }
    \label{fig:potential}
\end{figure*}

We study the gravitational potential of the individual haloes of the merger and the different components.
In Fig.~\ref{fig:potential}, we show the gravitational potential for the fSIDM equal-mass merger with $\sigma_\mathrm{\Tilde{T}}/m = 0.5 \, \mathrm{cm}^2 \, \mathrm{g}^{-1}$.
The gravitational potential is shown for the DMO (left-hand side) and the ICM (right-hand side) simulations at $\tau = 0.23$ (top row) and $\tau=0.64$ (bottom row).
We can see that the ICM stays closer to the barycentre and thus leads to a qualitative different total gravitational potential compared to the DMO setup. As a consequence, the peak positions are accelerated differently given the different potential gradients when comparing them at the same internal merger time $\tau$. This leads effectively to a tidal force acting on the offset between the DM and galaxy peaks which was initially created at about the first pericentre passage due to the drag force arising from the DM self-interactions. As a result, the offset at later times ($\tau \gtrsim 0.5$) is larger in the ICM runs than the ones in the DMO simulations.

\section{Temperature maps}
\begin{figure*}
    \centering
    \includegraphics[width=0.95\textwidth]{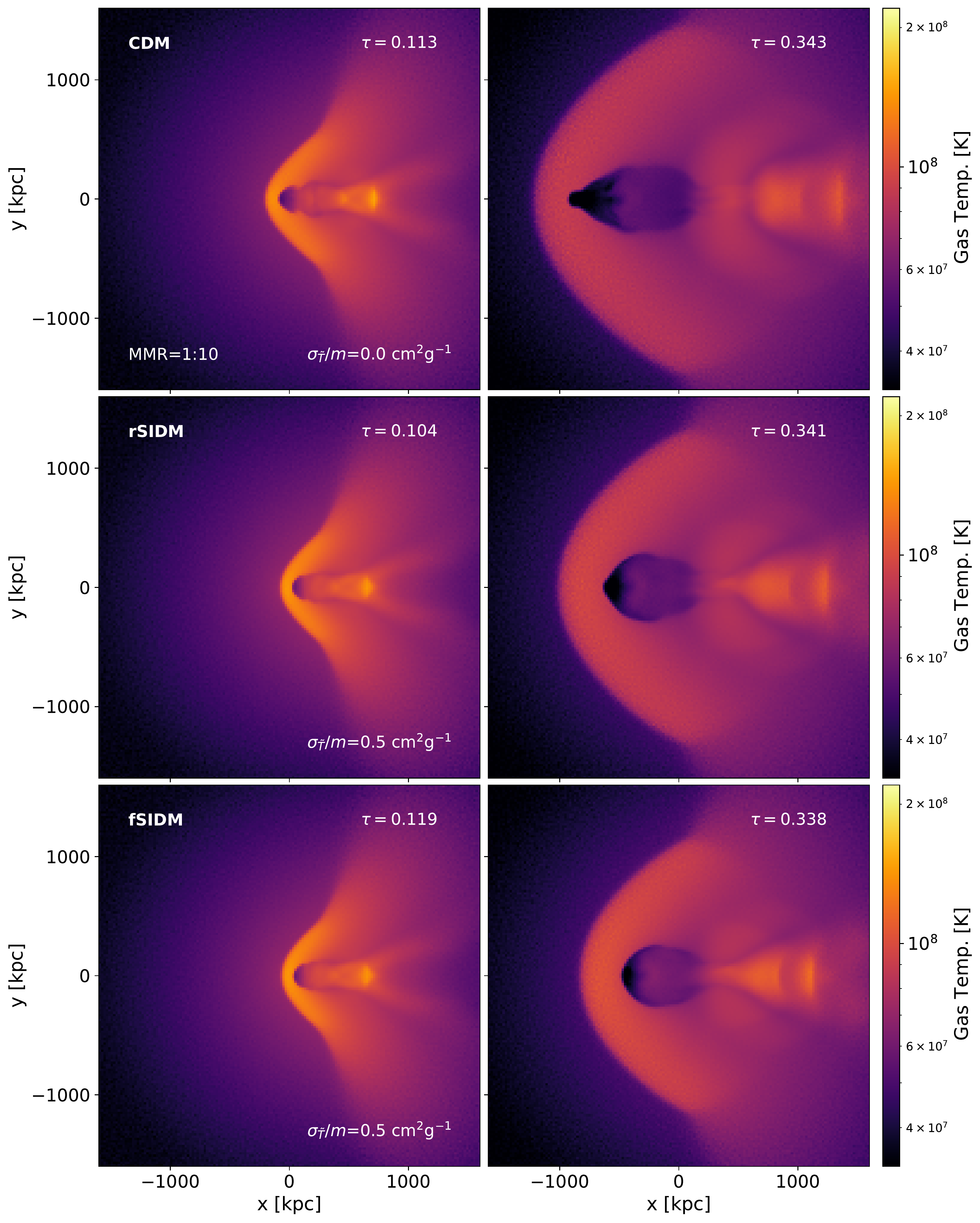}
    \caption{The mass-weighted temperature is shown in a projection perpendicular to the merger axis for the 1:10 mergers. Analogously to Fig.~\ref{fig:temp}, we show the results for CDM (top row), rSIDM (middle row), and fSIDM (bottom row). The temperature shortly after the first pericentre passage ($\tau \approx 0.11$) is shown on the left-hand side and a later stage ($\tau \approx 0.34$) is displayed on the right-hand side.}
    \label{fig:temp10}
\end{figure*}

In addition to Fig.~\ref{fig:temp}, we provide temperature maps for the 1:10 merger simulations in Fig.~\ref{fig:temp10}.
We compute them the same way as for the equal-mass merger, for details see Section~\ref{sec:morphology}.
The results are shown for two different times, $\tau \approx 0.11$ and $\tau \approx 0.34$.
For the earlier time, only small morphological differences are visible.
Whereas at later times, they have become more pronounced.
In particular, this concerns the shape of the cold gas region behind the left shock front.
Those morphological differences between DM models could eventually help to constrain the nature of DM.


\bsp	
\label{lastpage}
\end{document}